\documentstyle[epsf]{mn}
\newif\ifAMStwofonts

\newcommand{\etal}{et al.}

\newcommand{\jpb}{J. Phys. B: Atom. Mol. Op. Phys.\thinspace}
\newcommand{\mnras}{MNRAS\thinspace}
\newcommand{\apj}{ApJ\thinspace}

\newcommand{\aap}{A\&A\thinspace}

\ifoldfss
  \ifCUPmtlplainloaded \else
    \NewTextAlphabet{textbfit} {cmbxti10} {}
    \NewTextAlphabet{textbfss} {cmssbx10} {}
    \NewMathAlphabet{mathbfit} {cmbxti10} {} 
    \NewMathAlphabet{mathbfss} {cmssbx10} {} 
  \fi
  \ifAMStwofonts
    \ifCUPmtlplainloaded \else
      \NewSymbolFont{upmath} {eurm10}
      \NewSymbolFont{AMSa} {msam10}
      \NewMathSymbol{\upi}     {0}{upmath}{19}
      \NewMathSymbol{\umu}     {0}{upmath}{16}
      \NewMathSymbol{\upartial}{0}{upmath}{40}
      \NewMathSymbol{\leqslant}{3}{AMSa}{36}
      \NewMathSymbol{\geqslant}{3}{AMSa}{3E}

       \let\ge=\geqslant
    \fi
  \fi
\fi 

\ifnfssone
  \newmathalphabet{\mathit}
  \addtoversion{normal}{\mathit}{cmr}{m}{it}
  \addtoversion{bold}{\mathit}{cmr}{bx}{it}

  \newmathalphabet{\mathbfit} 
  \addtoversion{normal}{\mathbfit}{cmr}{bx}{it}
  \addtoversion{bold}{\mathbfit}{cmr}{bx}{it}
  \newmathalphabet{\mathbfss} 
  \addtoversion{normal}{\mathbfss}{cmss}{bx}{n}
  \addtoversion{bold}{\mathbfss}{cmss}{bx}{n}
  \ifAMStwofonts
    \ifCUPmtlplainloaded \else
      \UseAMStwoboldmath
      \makeatletter
      \new@mathgroup\upmath@group
      \define@mathgroup\mv@normal\upmath@group{eur}{m}{n}
      \define@mathgroup\mv@bold\upmath@group{eur}{b}{n}
      \edef\UPM{\hexnumber\upmath@group}
      \new@mathgroup\amsa@group
      \define@mathgroup\mv@normal\amsa@group{msa}{m}{n}
      \define@mathgroup\mv@bold\amsa@group{msa}{m}{n}
      \edef\AMSa{\hexnumber\amsa@group}
      \makeatother
      \mathchardef\upi="0\UPM19
      \mathchardef\umu="0\UPM16
      \mathchardef\upartial="0\UPM40
      \mathchardef\leqslant="3\AMSa36
      \mathchardef\geqslant="3\AMSa3E

       \let\ge=\geqslant
    \fi
  \fi
\fi 

\ifnfsstwo
  \DeclareMathAlphabet{\mathbfit}{OT1}{cmr}{bx}{it}
  \SetMathAlphabet\mathbfit{bold}{OT1}{cmr}{bx}{it}
  \DeclareMathAlphabet{\mathbfss}{OT1}{cmss}{bx}{n}
  \SetMathAlphabet\mathbfss{bold}{OT1}{cmss}{bx}{n}
  \ifAMStwofonts
    \ifCUPmtlplainloaded \else
      \DeclareSymbolFont{UPM}{U}{eur}{m}{n}
      \SetSymbolFont{UPM}{bold}{U}{eur}{b}{n}
      \DeclareSymbolFont{AMSa}{U}{msa}{m}{n}
      \DeclareMathSymbol{\upi}{0}{UPM}{"19}
      \DeclareMathSymbol{\umu}{0}{UPM}{"16}
      \DeclareMathSymbol{\upartial}{0}{UPM}{"40}
      \DeclareMathSymbol{\leqslant}{3}{AMSa}{"36}
      \DeclareMathSymbol{\geqslant}{3}{AMSa}{"3E}

       \let\ge=\geqslant
    \fi
  \fi
\fi 

\ifCUPmtlplainloaded \else
  \ifAMStwofonts \else 
    \def\upi{\pi}
    \def\umu{\mu}
    \def\upartial{\partial}
  \fi
\fi

\def\ti2{Ti~{\sc ii}} 
\def\ni2{Ni~{\sc ii}} 
\def\fe2{Fe~{\sc ii}}
\def\sr2{Sr~{\sc ii}}
\def\cm3{cm$^{-3}$} 
\def\etacar{$\eta$~Carinae\thinspace}

\title{[\ti2] and [\ni2] emission from the strontium filament of \etacar}
\author[M. Bautista et al.]{M.A.~Bautista$^1$,
H.~Hartman$^2$, T.R.~Gull$^3$, N.~Smith$^4$, K.~Lodders$^5$ \\
$^1$Centro de F\'{\i}sica, IVIC, P.O. Box 21827, Caracas 1020A, Venezuela\\ 
$^2$Lund Observatory, Lund University, Box 43, SE-22100 Lund, Sweden\\
$^3$Code 667, NASA Goddard Space Flight Center, Greenbelt MD, USA\\
$^4$Center for Astrophysics and Space Astronomy, University of Colorado,
389 UCB, Boulder, CO 80309, USA\\
$^5$Planetary Chemistry Laboratory, Dept. of Earth and Planetary Sciences,
Washington University, St. Louis, MO 63130, USA} 
\date{}
\pagerange{}
\pubyear{}

\begin{document}

\label{firstpage}

\maketitle

\begin{abstract}
We study the nature of the [\ti2] and [\ni2] emission from the so-called
{\it strontium filament} 
found in the ejecta of \etacar. To this purpose we
employ multilevel models of the \ti2 and \ni2 systems
which are used to investigate the physical condition of the filament and
the excitation mechanisms of the
observed lines.
For the \ti2 ion, for which no atomic data was previously available, we
carry out {\it ab initio} calculations of radiative transition
rates and electron impact excitation rate coefficients.
It is found that
the observed spectrum is consistent with the lines being 
excited in a mostly neutral
region with an 
electron density of the order of $10^7$ cm$^{-3}$ and a temperature
around  6000~K. In analyzing three observations with different
slit orientations recorded 
between March~2000 and November~2001 we find line ratios that 
change among various observations, in a way 
consistent with changes of up to an order of magnitude in the strength of the
continuum radiation field. These changes result from different 
samplings of the extended filament, due to the different slit
orientations used for each observation, and yield  clues on the spatial 
extent and optical depth of the filament. 
The observed emission indicates a large Ti/Ni abundance ratio
relative to solar abundances. 
It is suggested that the observed high Ti/Ni ratio in gas 
is caused by dust-gas fractionation processes and does not reflect the absolute Ti/Ni 
ratio in 
the ejecta of \etacar. 
We study the condensation chemistry of Ti, Ni and Fe within the
filament and suggest that the observed gas phase overabundance of Ti 
is likely the result of selective photo-evaporation of Ti-bearing
grains. Some mechanisms for such a scenario are proposed.
\end{abstract}


\section{Introduction} 

\etacar is the most luminous star in our part of the Galaxy
    and indeed is one of the most mysterious known to us (Davidson \&
    Humphreys 1997).  It is surrounded by complex ejecta created in
    several outbursts of the star and an outgoing massive wind with
periodic events every 5.54 years (Damineli 1996).  
The largest mass ejection occurred
    during the so-called Great Eruption observed in the 1840s -- which
    the star survived.  In that event, \etacar released about
    10$^{49.6}$ ergs of kinetic energy contained in more than 10
    M$_{\odot}$ of material (Smith et al.\ 2003), which has since
    expanded to form the bipolar Homunculus nebula we see today
    (Gaviola 1950; Smith \& Gehrz 1998; Morse et al.\ 2001).  Despite
    the hot and very luminous central star, the circumstellar nebula
    is cool enough to produce a remarkably rich emission spectrum of
    lines from neutral and low ionization species like Fe$^+$ and Ni$^+$ (Davidson
    et al.\ 2001; Zethson 2001; Zethson \etal 2001), line-of-sight absorption from several neutral and
    singly ionized atoms (Gull et al.\ 2005), and even emission from
    H$_2$ and very cool dust (Smith 2002a; Smith et al.\ 2003).  In
    addition to being dense and cool, the ejecta of \etacar are
    nitrogen rich, representing ashes of the CNO cycle (Davidson et
    al.\ 1982, 1986; Smith \& Morse 2004; Verner et al.\ 2005).  It is
    the high density, low-excitation, and CNO abundances of the ejecta
    that make the Homunculus a unique laboratory among circumstellar
    nebulae, having a profound effect on the results discussed in this
    paper.

In February 1999 a series of STIS observations of \etacar was made
of a region about 1.5" northwest of the star (Zethson \etal 2001). The spectrum 
from this region was significantly different from the nebular emission
spectra elsewhere in the Homunculus, 
with two of the lines being identified as forbidden emission of singly-ionized strontium ([\sr2]). 
The measured 
velocity of the filament ($\sim -100$ km/s)
and its distance from the central
star are consistent with the gas having been ejected from the same star
during the Great Eruption.

Further spectroscopic observations and a complete list of lines identified
in the filament were reported by Hartman et al. (2004). 
They show that lines of Fe~{\sc i}, V~II, and \ti2  are strong in the filament,
whereas only a few \fe2 lines  are detected. 
Other
lines in the spectrum at about the same velocity come from
C~{\sc i}, Mg~{\sc i}, Al~{\sc ii}, Ca~{\sc i}, Ca~{\sc ii}, Sc~{\sc ii}, 
Cr~{\sc ii}, Mn~{\sc ii}, Co~{\sc ii}, and \ni2. 
Curiously no H, He, S, O, or N lines are present in this region.

The cosmic abundance of strontium is lower than that of iron and nickel
by factors of 30,000 and 3,000 respectively. Thus, strontium emission
is normally undetectable in nebular spectra. The fact that [\sr2]   
emission was detected with strengths comparable to those of
[\fe2] and [\ni2] lines indicates that either strontium is locally
overabundant in \etacar by orders of magnitude, which is
inconsistent with the evidence of CNO-cycle-processes ejecta,  or  very 
unusual conditions exist in this region of the nebula. 

In a recent study of the \sr2  and [\sr2] spectra  of the filament
(Bautista \etal 2002) we found that [\sr2] emission may be
strongly enhanced by continuum fluorescence, which partially explains 
the strength of these lines. Although it was not possible at that 
time to establish the precise  abundance of Sr, this result did imply that 
Sr might not be as overabundant as previously thought.    
This point opened up new questions. If [\sr2]  lines can be
efficiently excited by fluorescence, why are the lines seen only 
in this structure, but not elsewhere in the Homunculus, and
not in any other emission line nebulae? 

The observed spatial  and velocity distribution of the Sr-filament suggests 
that it is located in the equatorial region between the hourglass-shaped 
bipolar Homunculus (Davidson et al 2001; Smith 2002a; Ishibashi et al, 2003).  
The Sr-filament is located at a projected distance of 10 lightdays (at 2300 pc 
distance), where it should receive intense UV radiation from \etacar. 
The spectra of the Weigelt Blobs B and D, located a few light days from \etacar,
require the spectral flux distribution of an O star with Te=35,000K during the 
5-year broad maximum (Verner \etal 2005). The Sr-filament should receive a 
diluted amount of this same UV radiation, but its emission line spectrum 
indicates a much lower ionization condition. The Sr-filament appears to 
be shielded not only by neutral hydrogen (13.6 eV), but also singly-ionized 
iron (7.9 eV) which requires a substantial column density of \fe2  
between \etacar and the Sr-filament. One possibility is that the 
Sr-filament is located at the outer side of an apparent 
optically thick, dusty, torus seen in the mid-IR (Smith et al. 2003).
In order to demonstrate this scenario, 
we must determine the nature of a hot star continuum filtered out by an 
\fe2  blocking column and predict the physical conditions within the filament 
that would sustain such a low ionization region. Then we can determine the actual abundances of all species observed in this structure.

In this paper we study the emission spectra of two prominent ions in the filament: \ti2  and \ni2. \ti2  is studied for the very first time as the relevant atomic data has never before been obtained. 
\ti2  emission lines are rarely seen in emission nebulae as its ionization potential, 13.58 eV, leads to Ti~{\sc iii} in most H~II regions. \ni2, often seen in nebular emission spectroscopy, 
is much better understood and herein serves as a reference ion.

The remainder of the paper is organized as follows: In Section 2
we present the atomic models built for the \ni2 and \ti2 systems,
with particular detail in the calculations of atomic parameters
that had to be performed for \ti2. In Section~3 we show the
results of the nebular diagnostics. In Section~4 we derive the
relative abundances of \ti2 and \ni2 in gas phase, while in Section~5
we discuss possible scenarios of Ni and Ti depletion in the nebula.
Finally, our discussion and conclusions are given in Section~6.

\section{Atomic data and atomic models for \ni2 and \ti2}

For the present analysis of [\ni2] spectra we employ a 76-level model
for the ion that includes continuum fluorescence excitation
as in Bautista \etal (1996) and improved collisional data reported by 
Bautista (2004).
This new collisional data, that results from very extensive calculations,
is expected to be more accurate than previous data.

\subsection{Radiative data for \ti2}

For \ti2, neither complete transitions rates for forbidden
transitions nor collision strengths had been reported, thus large calculations
of atomic data for this ion were carried out within the present work.

\begin{table}          
\caption{\ti2 configurations and scaling parameters} 
\centering
\begin{tabular}{l}       
 \hline     \hline
Spectroscopic configurations\\ 
$3s^2\ 3p^6\ 3d^3$, $3s^2\ 3p^6\ 3d^2\ 4s$, $3s^2\ 3p^6\ 3d^2\ 4p$, $3s^2\ 3p^6\ 3d\ 4s^2$\cr 
\hline
Correlation configurations \cr

$3s^2\ 3p^6\ 3d^2\ \bar{4d}$, 
$3s^2\ 3p^6\ 3d\ 4s\ 4p$, 
$3s^2\ 3p^6\ 3d\ 4s\ \bar{4d}$,
\cr
$3s^2\ 3p^6\ 3d\ 4s\ \bar{5s}$, 
$3s^2\ 3p^6\ 3d\ 4s\ \bar{5d}$, 
$3s^2\ 3p^6\ 3d\ 4s\ \bar{5p}$,
\cr
$3s^2\ 3p^6\ 3d\ 4s\ \bar{4f}$, 
$3s^2\ 3p^6\ 3d\ 4p^2$, 
$3s^2 3p^5\ 3d^4$,
$3s^2\ 3p^5\ 3d^3\ 4s$, 
\cr
$3s^2\ 3p^5\ 3d^3\ 4p$, 
$3s^2\ 3p^5\ 3d^2\ 4s^2$, 
$3s^2\ 3p^4\ 3d^5$, 
$3s^2\ 3p^4\ 3d^3\ 4s^2$\cr
$3s^2\ 3p^4\ 3d^4\ 4s$,
$3s^2\ 3p^4\ 3d^4\ 4p$,
$3s\ 3p^6\ 3d^4$,
$3s\ 3p^6\ 3d^3\ 4s$,
\cr
$3s\ 3p^6\ 3d^3\ 4p$
$3s\ 3p^5\ 3d^4\ 4s$,
$3s\ 3p^5\ 3d^4\ 4p$,
$3p^6\ 3d^4\ 4s$,
\cr
$3p^6\ 3d^4\ 4p$\cr
\hline
$\lambda_{nl}$\cr
1.43570(1s), 1.12290(2s), 1.06650(2p), 1.09690(3s),
1.08430(3p),\cr
1.12590(3d), 1.27470(4s), 1.22660(4p), -0.79940($\bar{4d}$) \cr
-0.64020($\bar{5s}$), 2.99520($\bar{5d}$), 1.90030($\bar{5p}$), 1.88930($\bar{4f}$)\cr
\hline
\end{tabular}
\end{table}

\begin{table}          
\caption{Energy terms in Rydberg for the Ti~II ion
relative to the ground state. The table compares experimental
energies with {\it ab initio} calculated
energies (Theo.) 
and energies with term energy corrections (Theo.-TEC)}    
\centering
\begin{tabular}{rllrrr}
 \hline     \hline
  & Config. & Term& Theo. & Theo.-TEC & Exp. \\
            \hline
  1  &$3d^2(^3F)4s       $&$ a~^4F  $&0.0     & 0.00    &0.0      \\ 
  2  &$3d^3             $&$ b~^4F  $&0.08806 & 0.00834 & 0.007839\\
  3  &$3d^2(^3F)4s       $&$ a~^2F  $&0.06439 & 0.04231 & 0.041528\\
  4  &$3d^2(^1D)4s       $&$ a~^2D  $&0.08922 & 0.05534 & 0.077508\\
  5  &$3d^3             $&$ a~^2G  $&0.09982 & 0.08140 & 0.080552\\
  6  &$3d^3             $&$ a~^4P  $&0.09202 & 0.08543 & 0.084077\\ 
  7  &$3d^3             $&$ a~^2P  $&0.09217 & 0.07588 & 0.088476\\
  8  &$3d^2(^3P)4s       $&$ b~^4P  $&0.10519 & 0.09021 & 0.088784\\
  9  &$3d^3             $&$ b~^2D  $&0.13436 & 0.11686 & 0.113737\\ 
 10  &$3d^3             $&$ a~^2H  $&0.14976 & 0.11788 & 0.113935\\
 11  &$3d^2(^1G)4s       $&$ b~^2G  $&0.18304 & 0.14025 & 0.137017\\ 
 12  &$3d^2(^3P)4s       $&$ b~^2P  $&0.19358 & 0.16370 & 0.149115\\ 
 13  &$3d^3             $&$ b~^2F  $&0.22551 & 0.19290 & 0.188561\\
 14  &$3d4s^2           $&$ c~^2D  $&0.37419 & 0.20894 & 0.226677\\
 15  &$3d^2(^3F)4p       $&$ z~^4G^o$&0.24771 & 0.26968 & 0.270750\\
 16  &$3d^2(^3F)4p       $&$ z~^4F^o$&0.25782 & 0.28059 & 0.281422\\   
 17  &$3d^2(^3F)4p       $&$ z~^2F^o$&0.26630 & 0.27996 & 0.285498\\ 
 19  &$3d^2(^1S)4s       $&$ a~^2S  $&0.35738 & 0.28071 & 0.287619\\
 18  &$3d^2(^3F)4p       $&$ z~^2D^o$&0.27074 & 0.29066 & 0.288805\\ 
 20  &$3d^3             $&$ d~^2D1 $&0.28882 & 0.34006 & 0.292376\\ 
 21  &$3d^2(^3F)4p       $&$ z~^4D^o$&0.27443 & 0.29206 & 0.295841\\ 
 22  &$3d^2(^3F)4p       $&$ z~^2G^o$&0.30665 & 0.31108 & 0.313768\\ 
 23  &$3d^2(^3P)4p       $&$ z~^2S^o$&0.32376 & 0.33886 & 0.339041\\ 
 24  &$3d^2(^1D)4p       $&$ z~^2P^o$&0.34940 & 0.35509 & 0.356748\\ 
 25  &$3d^2(^1D)4p       $&$ y~^2D^o$&0.35800 & 0.35671 & 0.358146\\ 
 26  &$3d^2(^1D)4p       $&$ y~^2F^o$&0.35890 & 0.35852 & 0.362557\\
 27  &$3d^2(^3P)4p       $&$ z~^4S^o$&0.35195 & 0.36370 & 0.362702\\  
 28  &$3d^2(^3P)4p       $&$ y~^4D^o$&0.36084 & 0.36767 & 0.368031\\
 29  &$3d^2(^3P)4p       $&$ z~^4P^o$&0.37814 & 0.38047 & 0.381834\\
 30  &$3d^2(^1G)4p       $&$ y~^2G^o$&0.40620 & 0.39781 & 0.396745\\
 31  &$3d^2(^3P)4p       $&$ x~^2D^o$&0.41916 & 0.41092 & 0.407173\\ 
 32  &$3d^2(^3P)4p       $&$ y~^2P^o$&0.42858 & 0.40434 & 0.412787\\ 
 33  &$3d^2(^1G)4p       $&$ z~^2H^o$&0.43455 & 0.41952 & 0.415324\\ 
 34  &$3d^2(^1G)4p       $&$ x~^2F^o$&0.45445 & 0.43599 & 0.431113\\
 35  &$3d^2(^1S)4p       $&$ ^2P^o  $&0.61270 & 0.57783 & 0.569851\\
            \hline
         \end{tabular}
\end{table}

For the calculation of radiative transition probabilities 
and for the wavefunctions of the \ti2 system 
we use the atomic structure code AUTOSTRUCTURE (Badnell 1986) to reproduce the structure of
the ion. This code is based on the program SUPERSTRUCTURE originally
developed by Eissner et al. (1974). In this approach the wavefunctions
are written as a configuration interaction expansion of the type:
\begin{equation}
\psi_i = \sum_j \phi_jc_{ji}
 \label{cero}
\end{equation}
where the coefficients $c_{ji}$ are chosen so as to diagonalize 
$\langle \psi_i \mid H \mid \psi_j \rangle$, where H is the
 Hamiltonian and the  basic functions $\phi_j$ are constructed from
one-electron orbitals generated using the Thomas-Fermi-Dirac model
potential (Eissner \& Nussbaumer 1969). The $\lambda_{nl}$ scaling
parameters in this model are optimized by minimizing a weighted sum of energies.
Representing a  system as complex as \ti2
requires many configurations, which must be selected rather carefully as to
account for all strongly coupled configurations without exceeding 
available computational resources. Experience shows that configurations with 
one and two electron promotions out of the $3s$, $3p$, and $3d$ orbitals 
are often important.
Further, improvements can be obtained in some cases through the
use high laying non-physical orbitals.
For the present \ti2 model it was necessary to 
include several non-physical orbitals in the expansion, i.e. $\bar{4d}$, $\bar{5d}$, $\bar{5s}$,
$\bar{5p}$, $\bar{4f}$. 
The configuration representation and scaling parameters used in this work
are listed in Table~1.

Fine tuning (semi-empirical corrections)---which is useful for
treating states that decay through
weak relativistic couplings (e.g. intercombination transitions)---takes the
form of term energy corrections (TEC). By considering the
relativistic wavefuntion, $\psi^{\rm r}_i$, in a perturbation expansion
of the non-relativistic functions $\psi^{\rm nr}_i$,
\begin{equation}
\psi^{\rm r}_i=\psi^{\rm nr}_i +\sum_{j\neq i}\psi^{\rm nr}_j\times
{\langle\psi^{\rm nr}_j|H_{\rm 1b}+H_{\rm 2b}|\psi^{\rm nr}_i\rangle\over
E^{\rm nr}_i-E^{\rm nr}_j}\ ,
\end{equation}
where $H_{\rm 1b}$ and $H_{\rm 2b}$ are the one- and two-body Breit-Pauli
relativistic operators. Thus, 
a modified $H_{nr}$ is constructed with improved estimates of the differences
$E^{\rm nr}_i-E^{\rm nr}_j$ so as to adjust the centers of gravity of the
spectroscopic terms
to the experimental values.
Table~2 shows a
comparison between the 
observed energies averaged over the fine structure
taken from Huldt et al. (1982), Moore (1949), and Russell (1927)
and compiled in  the NIST energy levels database (NIST 2000), the 
computed {\it ab initio} term energies, and the computed energies
after TEC corrections.
The overall agreement between {\it ab initio} energies and experimental
values is typically within 10\%, with better agreement for the 
odd parity terms but with some problems for low laying even parity multiplets. 
To the limit of our computations, we were unable to find a set of 
correlation 
configurations able to reproduce {\it ab initio} the correct order of
the terms $3d^3\ a~^2G,\ a~^4P$, and $a~^2P$ 
(terms 5 through 7). We also found it difficult to reproduce the position
and an accurate energy for the $3d4s^2~c~^2D$ term. These 
large discrepancies between calculated and experimental energies were readily 
resolved by the TEC technique, although some corrections
were uncomfortably large. 
A noticeable exception is the $a~^2D$ term which lowers its energy far from
the experimental value after TEC corrections. This term could not be improved 
further as it has a profound effect on the perturbative corrections on
the $b~^4F$ and $a~^2F$ terms.

This ion's representation was used to compute radiative transition rates
for both dipole-allowed and forbidden transitions. Table~3 shows a comparison 
between the present $log(gf)$-values for dipole allowed transitions and those
determined experimentally by Pickering et al. (2001, 2002) 
and Bizzarri et al. (1993), 
computed
by Kurucz (2000), and those compiled in the NIST database (taken from
Meggers et al. 1975; Roberts et al. 1975; Danzmann and Kock 1980; 
Blackwell et al. 1982; Roberts et al. 1973). Most $gf$-values
in the NIST database have an accuracy rating "D", which means that the 
uncertainties are greater than 50\%. Table~3 is limited to 
lines with $gf$-values greater than 0.01 for which reasonably 
accurate data is expected
given the present representation of the ion. We find good agreement,
within 0.2~dex, 
between our results and those of other authors for $log(gf)> -1$
and a mean dispersion of about 0.5~dex for smaller values. 
Given the uncertainties in the present results, for our
\ti2 multilevel model we combine the NIST $f$-values
whenever available with the present radiative data for all other
transitions. 

\begin{table}
 \centering
\caption{Comparison between  $log(gf)$-values for dipole allowed transitions 
from the present work, and those of Pickering et al. (2001, 2002),
Bizzarri et al. (1993), Kurucz (2000), and the values recommended
by NIST}  
\begin{tabular}{cccccc}
\hline\hline
  Wave. (nm) & Present & Pickering & Bizzarri & Kurucz& NIST  \\
\hline
\multicolumn{6}{l}{Upper level: 29544 cm$^{-1}$ $3d^2(^3F)4p~z~^4G_{5/2}$}\\
\hline
 401.23835 & -1.14  & -1.84 & -1.75 & -1.73  & -1.61\\
 349.10494 & -1.01  & -1.15 & -1.06 & -1.13  &  \\
 340.98043 & -1.80  & -1.98 & -1.89 & -1.89  & -1.90 \\
 339.45720 & -0.56  & -0.55 & -0.54 & -0.52  & -0.59 \\
 338.37587 & -0.052  &  0.16 &  0.15 &  0.20  & 0.050 \\
\hline
 & & & & \\
\multicolumn{6}{l}{Upper level: 31490 cm$^{-1}$ $3d^2(^3F)4p~z~^2F_{7/2}$}\\
\hline 
 465.72004 & -2.68  & -1.59 & -1.63 & -1.59  & -2.15 \\ 
 439.50311 & -0.92  & -0.54 & -0.51 & -0.45  & -0.66 \\
 375.92914 &  0.16  &  0.28 &  0.27 &  0.31  & 0.20 \\
 321.47669 & -3.30  & -1.40 & -1.34 & -1.35  & -1.39 \\
 319.75185 & -0.30  & -1.97 & -1.90 & -2.55  & -1.87 \\
 1587.3644 & -1.38  &       &       & -1.92  &     \\ 
\hline
\end{tabular}
\end{table} 

Our representation of the ion was also used to compute electric
quadrupole and magnetic dipole transition rates. The results of such 
calculations are rather sensitive to configuration interactions and
level mixing. Some indication about the accuracy of the present 
rates is obtained from recent measurements by the FERRUM project
(Hartman et al. 2003, 2005) of lifetimes of metastable levels,
which decay through parity forbidden transitions. Lifetimes 
for five levels have been reported so far, which are compared
in Table~4 with our theoretical results. 
Our calculated lifetimes agree within $\sim$50\% with the 
experimental determinations, which is the estimated overall 
uncertainties for the present radiative rates for forbidden
transitions. 

\begin{table}
 \centering
\caption{Lifetimes (in sec) for metastable levels of \ti2 as determined
experimentally by Hartman  et al. (2003, 2005) and from the
present calculations}
\begin{tabular}{cccccc}
\hline\hline
  Level 
        &  Hartman & present \\
            \hline\noalign{\smallskip}
 $3d^2(^3P)4s~b~^4P_{3/2}$   & 18(4)  & 31.1    \\
 $3d^2(^3P)4s~b~^4P_{5/2}$   & 28(10)  & 16.4    \\ 
 $3d^2(^3P)4s~b~^2P_{1/2}$   & 14(3)  & 12.7    \\
 $3d4s^2~c~^2D_{3/2}$ & 0.29(0.01)  & 0.192   \\ 
 $3d4s^2~c~^2D_{5/2}$ & 0.33(0.02)  & 0.196 \\
\hline 
\end{tabular}
\end{table}

\subsection{Collisional data for \ti2}

We have computed electron impact 
collision strengths for transitions among the lowest 82 energy levels
of \ti2 that belong to the configurations 
$3d^3$, $3d^24s$, $3d^24p$, and 
$3d4s^2$. 
We employ the Breit-Pauli R-matrix method and the BPRM set of
codes developed in the framework of the IRON Project (Hummer \etal  1993;
Berrington et al. 1995).
Partial wave contributions are included from 60 $SL\pi$
total symmetries with angular momentum $L = 0 - 9$, total multiplicities 
$(2S+1) = 1 - 5 $, and parities even and odd.
Then, the collision strengths were "topped up" with estimates of contributions from higher partial waves
based on the Coulomb-Bethe approximation (Burgess 1974).
Collision strengths for the fine-structure levels were obtained by algebraic
recoupling
of the LS reactance matrices (Hummer \etal  1993). 
The collision strengths were calculated at $11000$ 
energy points from 0 to 6 Ry, 
with 90\% of the points in
the region with resonances. This number of  energy points was found to
provide sufficiently good resolution for accurate calculations of 
the contributions of autoionizing resonances to the effective collision 
strengths at temperatures between  5000~K and 20000~K.
Figure 1 shows the collision strengths for some of the dominant transitions. 

\begin{figure*}
  \leavevmode
    \epsfysize=18cm
\epsfbox{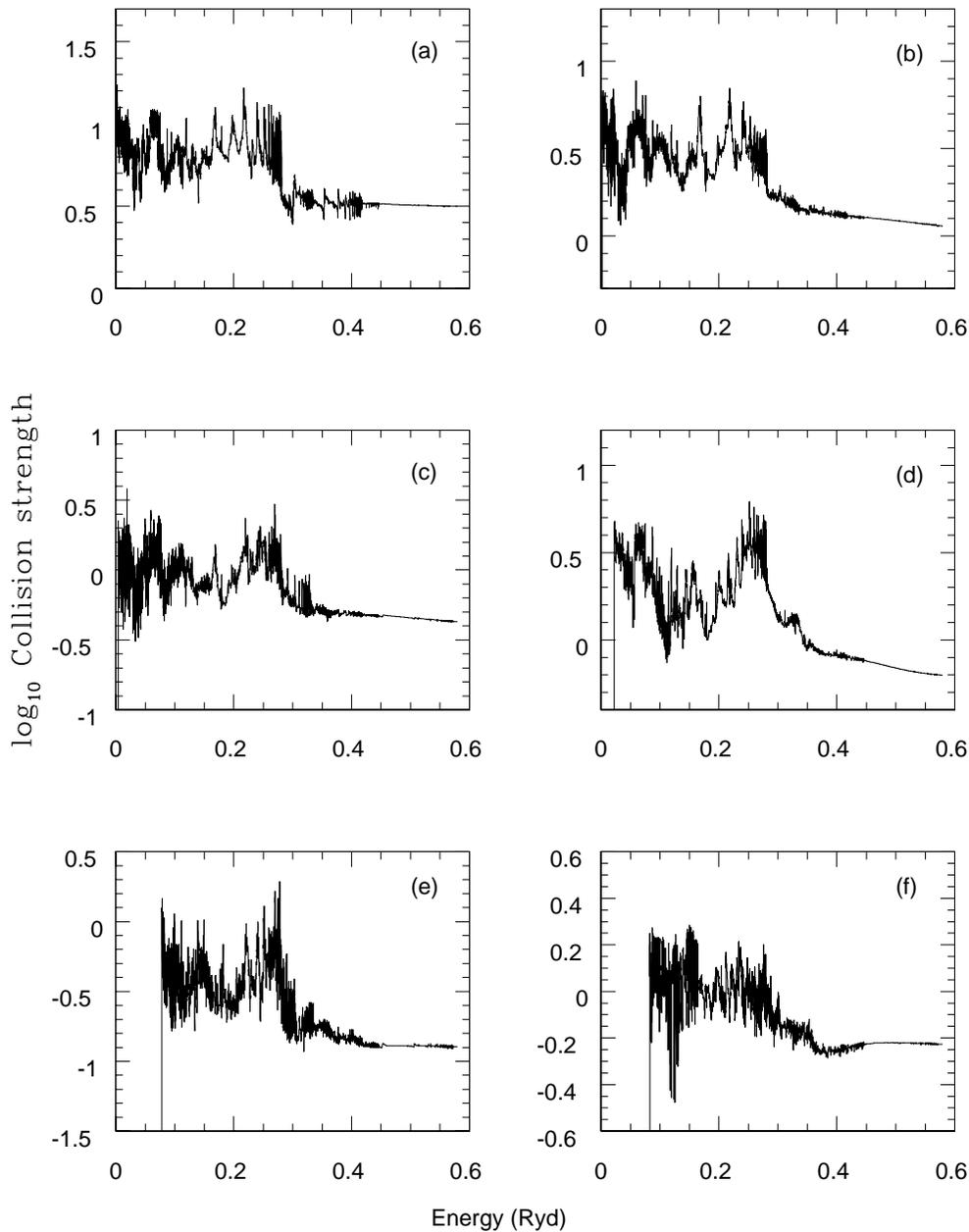}
  \caption{Collision strengths vs. incident electron energy for excitations from the ground level $a~^4F_{3/2}$ to $a~^4F_{5/2}$ (a),
$a^4F_{7/2}$ (b), $a~^4F_{9/2}$ (c), $b~^4F_{3/2}$ (d), $a~^2F_{7/2}$ (d), $a~^2D_{3/2}$ (e).} 
  \label{fig1}
\end{figure*}

We compute Maxwellian-averaged effective collision strengths, 
defined as:
\begin{equation}
\Upsilon(T) = \int_0^{\infty} \Omega_{if}(\epsilon_f)\exp(-\epsilon_f/kT)~d(\epsilon_f/kT),
\end{equation}
where $\Omega_{if}$ is the collision strength for the transition $i$ to $f$, and 
$\epsilon_f$ is the energy of the outgoing electron. The $\Upsilon$'s were computed for 
various temperatures between 5000 and 20~000K to be used in the calculation of excitation rates.

\subsection{Multilevel model of \ti2}    

We built an 82-level
excitation equilibrium model for \ti2   
which considers both electron impact excitation and continuum fluorescence excitation. 
A model of this size is large enough to account for collisionally excited emission
at the temperatures of interest ($\sim 10^4$ K), 
and the dominant 
photoexcitation channels from the ground level. If more
levels were included in our models they would tend to increase the
efficiency of 
continuum fluorescence excitation, although it has been shown that 
the contribution of individual levels decreases rapidly with the level
energy (Bautista et al. 1996).
We assume that the radiation field density, $U_\nu$, at photon energies
below the Lyman ionization limit (13.6 eV) can be approximated
by a blackbody with temperature $T_R$ times a geometrical dilution
factor $w$, i.e.
$$
{c^3U_\nu\over 8\pi h \nu^3} = {w\over \exp(h\nu/kT_R) - 1}
$$
For the present calculations we adopt a blackbody temperature of 35~000K
(see Verner \etal 2005) 
while the dilution factor is diagnosed from the observed spectra.
A different choice in the blackbody temperature would lead to
a change in the diagnosed dilution factor, but would have
negligible effect on the modeled spectrum.

Fig.~2 presents a diagram 
of the dominant      
photo-excitation channels for two [\ti2] lines arising from
quartets and doublets terms. The main photoexcitation 
channels are indicated by dashed lines and the radiative 
deexcitation transitions are
marked by solid lines. Such dipole allowed radiative decays
lead to emission lines, observable in the spectra,
just as they populate the levels that give rise
to forbidden lines.
It is worth noticing that in the absence of collisional excitation,
the strength of forbidden lines would be tightly bound to the
intensity of the dipole transitions from the pumping mechanism.
On the other hand, the contribution of collisional excitations
to the forbidden lines introduces a complex temperature and density 
dependence in the relative strengths of dipole and forbidden 
transitions, which can be used for spectral diagnostics as illustrated
in the next section.

Fig.~3 presents the line emissivity per ion of the 
 \ti2 lines [$\lambda6648$] 
($b~^2G_{7/2}~-~a~^4F_{7/2}$) and [$\lambda4918$] ($c~^2D_{3/2}~-~a~^2F_{5/2}$)
and the blend [$\lambda10069b$]   
($b~^4P_{3/2}~-~a~^4F_{3/2}$ + $b~^4P_{5/2}~-~a~^4F_{5/2}$) 
vs. $N_e$ for an electron temperature of 6000~K, 
a blackbody temperature of 35~000~K and
four values of the dilution factor,
$w=$ 0, $10^{-10}$, $5\times10^{-10}$, and $10^{-9}$. For 
electron densities less than $N_e\sim10^7$\cm3 continuum fluorescence 
dominates the strength of most lines. On the other hand,
for electron densities around $N_e\sim10^7$, as previously diagnosed for the
strontium filament (Bautista et al. 2002), collisional excitation
dominates the 
[$\lambda6648$] and [$\lambda10069b$] features, while fluorescence 
still makes a very significant contribution to the
excitation of the $c~^2D_{3/2}$ level, responsible for the
[$\lambda4918$] line. This is 
because collisional transition  rates from the ground levels 
to the $c~^2D$ multiplet are
rather small.

\begin{figure}
  \leavevmode
    \epsfysize=11cm
    \epsfbox{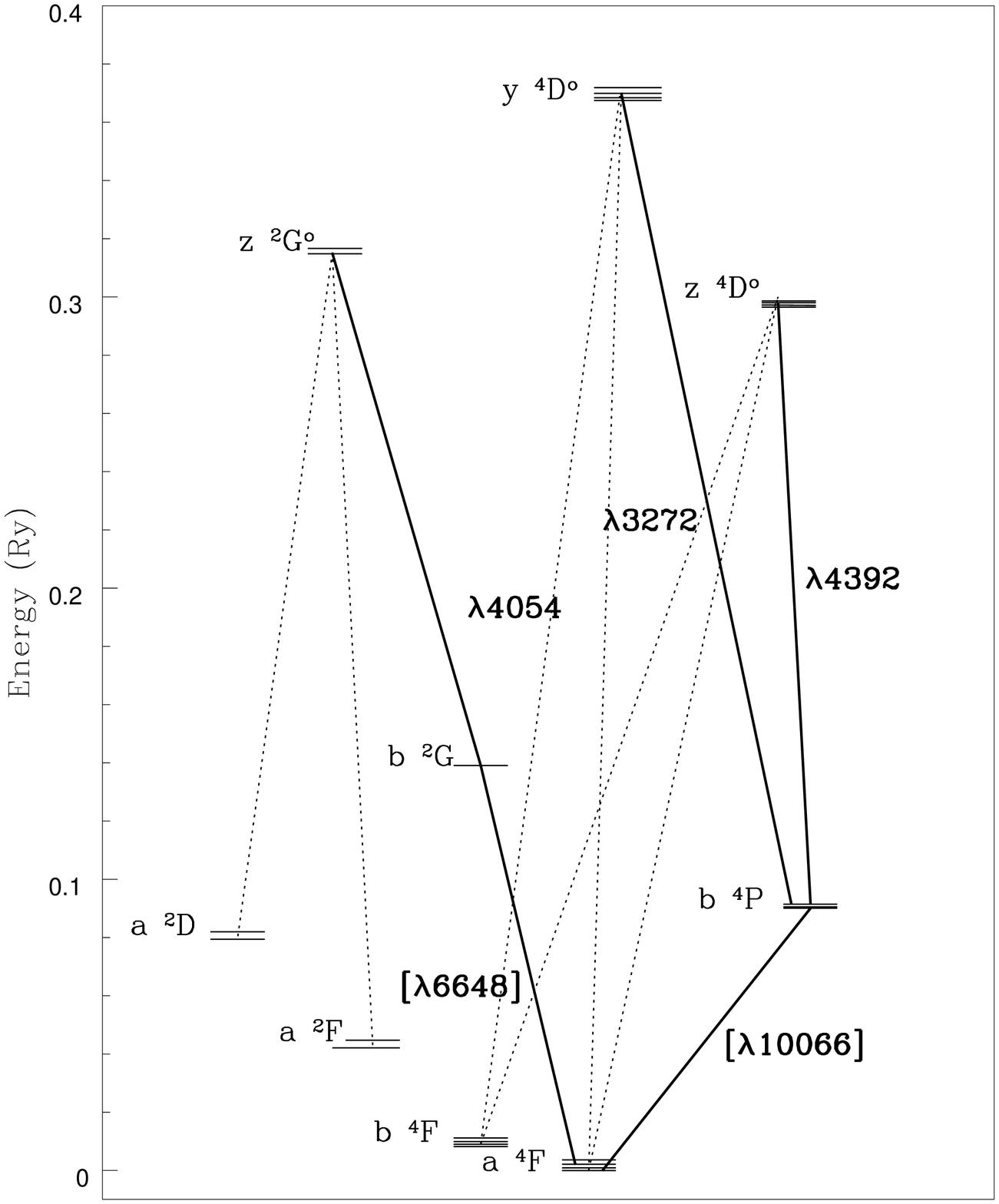}
  \caption{Dominant photoexcitation channels for two [\ti2] lines.
Some of the dipole allowed lines also emitted in the photo-pumping
process are also indicated}
  \label{fig2}
\end{figure}

\begin{figure}
  \leavevmode 
\epsfysize=12cm
\epsfbox{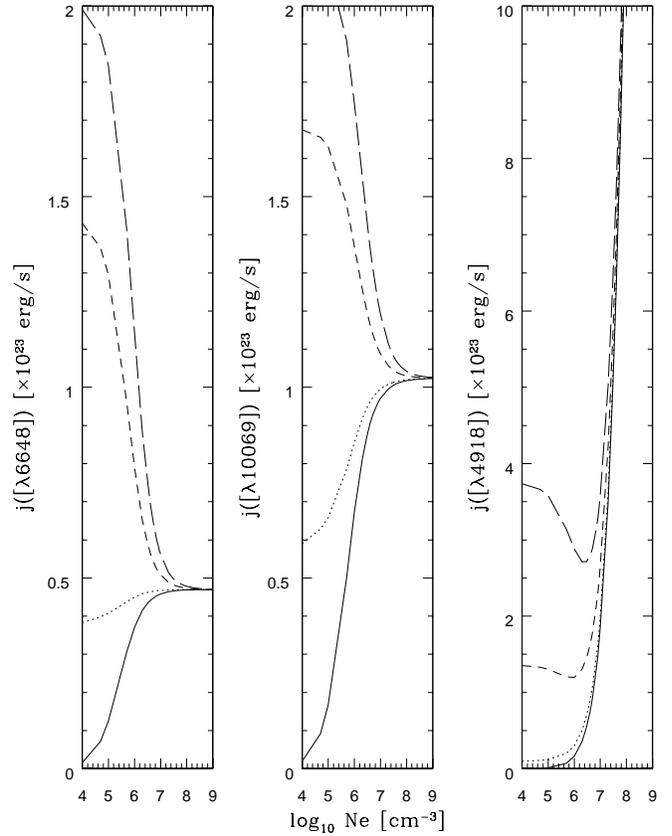}
  \caption{Line emissivity of the
[\ti2] $\lambda6648$, $\lambda 10069$ and $\lambda4918$ lines ion vs. electron density
for dilution factors, $w=$ 0 (solid line), $10^{-10}$ (dotted line),
$5\times10^{-10}$ (short-dashed line), and $10^{-9}$ (long-dashed line).
The curves are calculated for $T_e=6000~K$, $N_e=10^7~cm^{-3}$, and
a blackbody temperature of 35~000~K.}
  \label{fig3}
\end{figure}

\section{Spectral diagnostics}

We use the [\ni2] and [\ti2] emission line strengths 
 to diagnose the physical conditions of 
the filament on each of the three HST/STIS observations:
March~2000, April~2001, and November~2001 (hereafter mar00, apr01, nov01
respectively; Bautista \etal 2002;
Hartman \etal 2004). Previous diagnostics
were reported in Bautista et al. (2002) from the  [Sr~{\sc ii}] lines.
There we estimated an electron density of the order of $10^7$ \cm3
and an upper limit to the dilution factor for photoexcitation
of $\sim 10^{-9}$.

\subsection{Observed line intensities}

For the present work, we use the deep spectra extending from 1640\AA\ 
to 10100\AA\ 
analyzed by Hartman et al. (2004). The original data was recorded during
three visits with the Hubble Space Telescope/Space Telescope Imaging Spectrograph
(HST/STIS) under several different observational programs. For the mar00
observation 
the long 52"x0.2" aperture was aligned close to the axis of the Homunculus
(Position Angle -41deg), whereas the aperture was aligned at 22~deg in apr01
and -130~deg in nov01. The different angles were
defined primarily by HST sun-angle restrictions, but were used to study
the extent of the Strontium filament in different directions. The intersection
of these apertures (see Fig.~1 in Hartman \etal 2004) is on the apparent peak
of the Strontium filament as first noticed by the discovery observations
(Zethson \etal 2001).

The line fluxes were integrated across the line profiles of the components
centered at $\sim$-100 km/s at specific positions along the slit. The intersection
position of the different visits is the primary position of interest. However
as noted below, several additional positions were sampled along the aperture 
oriented radially to \etacar (P.A.=-41$^o$) and tangentially (-130$^o$) to
understand the spatial variation of specific line ratios and therein, to 
study the excitation and ionization structure of the filament.
The measured line intensities vary along the aperture and between visits.
The structure is complex as other nebular structures are in line of sight:
namely the Homunculus and the Little Homunculus. At some positions, Hartman
\etal (2004) noted a second velocity structure in many lines. These two velocity systems are likely related to the two disks noted by Davidson \etal (2001)
located between the bipolar lobes. 
The two velocity components correspond to material in the same
   equatorial disk but with two different ages, as noted by Davidson
   et al. (2001).  The more prominent feature is associated with the parts
   of the skirt ejected along with the Homunculus in the 1840's.  The
   fainter and more blueshifted feature is also in the equator, but was
   ejected in the 1890 outburst along with the Little Homunculus
   (Ishibashi et al. 2003; Smith 2005) and the Weigelt knots (Smith et
   al. 2004; Weigelt et al. 1995).  In this study we focus on the
   structure ejected in the 1840s.

As with any spectroscopic analysis the observed line intensities must be  
corrected for extinction. None of the lines that form the classical line
ratios
(e.g. hydrogen recombination lines) are detected. We use two relatively well
known [\ni2] transitions that arise from a common upper level,
i.e. $\lambda 7413$$(a~^2D_{3/2}~-~a~^2F_{5/2}$) and $\lambda 6668$$(a~^2D_{5/2}~-~a~^2F_{5/2}$), and the standard
interstellar extinction curve of Cardelli \etal (1989). 
While these lines are relatively close in  wavelength, thus far they are
the best candidates for estimating extinction as all other lines of interest are much weaker or their predicted line ratios are less certain.
The observed line ratios are $2.20\pm0.13$ in mar00, 
$1.94\pm0.13$ in apr01,
and $2.07\pm0.13$ in nov01. Thus, there are no significant variations
of the line ratio between the three observations.
From these observations we derive 
an extinction magnitude, $A_V=2.1\pm0.9$, as  
required to correct the ratios to the theoretical value
of 1.6, with an estimated uncertainty of $\sim$10\%.
The relatively large uncertainty in $A_V$ comes from the fact that the
two lines used in the diagnostic are close in wavelength. 
This value for the extinction magnitude is consistent with the value
measured, $A_V=2.06$, from photometry of stars of the Tr~16 cluster
in the Carinae nebula 
(Tapia \etal 2003). The actual extinction magnitude to the filament could be 
somewhat higher than to the Carinae stars due to the dust within
the nebula. 
Smith (2002b) studied local extinction within the larger Carina Nebula,
  and found a total-to-selective extinction of
  R=$A_V/E(B-V)=4.8$, which is different than the normal ISM value of
  3.1.  Furthermore, the extinction around \etacar    itself is likely
  to be very gray because of the large grain size (Smith et al. 2003).
  However, thermal-IR maps of the dust column density in the Homunculus
  reveal a local minimum near the position of the Sr filament (Smith et
  al. 2003), so the unusual reddening there may not be a severe problem.
We thus adopt for the remaining of this paper a extinction
law with $A_V=2.1$ and the interstellar $R=3.1$. Clearly, the remaining
uncertainty on this value can be important in trying to compare lines
well separated in wavelength, thus all diagnostics should be carried out with
lines 
close to each other. 

Tables~5 and 6 present the complete lists of [\ni2] and [\ti2] lines
used for the present study and their intensities corrected for
extinction.

\begin{table*}
 \centering
\caption{[\ni2] lines from the Sr-filament} 
\begin{tabular}{lcrrrrrr}
\hline\hline
$\lambda_{lab}$ &Transition
&\multicolumn{3}{c}{Observed flux ($10^{-13}$} & 
\multicolumn{3}{c}{Extinction corrected flux} \\
 & & \multicolumn{3}{c}{ergs sec$^{-1}$ cm$^{-2}$ arcsec$^{-1}$)}
& \multicolumn{3}{c}{($A_V=2.1$)} \\
            \hline\noalign{\smallskip}
 & & Mar00 & Apr01 & Nov01 & Mar00 & Apr01 & Nov01 \cr
\hline
7379.970&$a~^2F_{7/2} - a~^2D_{1/2}$&286.00 &211.00 &239.00 &1088.0 &802.7  &909.2  \cr
8303.270&$a~^2F_{7/2} - a~^2D_{3/2}$&       &       &11.80  &       &       &34.6   \cr
        &                           &       &       &       &       &&      \cr      
6668.670&$a~^2F_{5/2} - a~^2D_{1/2}$& 64.60 &41.60  &51.30  &304.3  &195.9  &241.6  \cr
7413.690&$a~^2F_{5/2} - a~^2D_{3/2}$&142.0 &80.60  &106.0   &534.8  &303.5  &399.2  \cr
        &                           &       &       &       &       &&      \cr       
4327.453&$a~^4P_{5/2} - a~^2D_{1/2}$&       &18.10  &27.00  &       &246.8  &368.1  \cr
7257.860&$a~^4P_{5/2} - a~^4F_{7/2}$& 23.00 &12.80  &10.70  & 90.8  &50.5   &42.2   \cr
        &                           &       &       &       &       &&      \cr     
6815.450&$a~^4P_{3/2} - a~^4F_{5/2}$& 15.00 &5.71   &7.32   & 67.6  & 25.7  &33.0   \cr
        &                           &       &       &       &       &       &      \cr
6793.350&$a~^4P_{1/2} - a~^4F_{5/2}$& 10.40 &3.50   &19.70  & 47.2  & 15.9  &89.4   \cr
      & & & & \cr
\hline
\end{tabular}
\end{table*}

\begin{table*}
 \centering
\caption{[\ti2] lines from the Sr-filament}  
\begin{tabular}{lcrrrrrr}
\hline\hline 
$\lambda_{lab}$ &Transition
&\multicolumn{3}{c}{Observed flux ($10^{-13}$} & 
\multicolumn{3}{c}{Extinction corrected flux} \\
 & & \multicolumn{3}{c}{ergs sec$^{-1}$ cm$^{-2}$ arcsec$^{-1}$)}
& \multicolumn{3}{c}{($A_V=2.1$)} \\
            \hline\noalign{\smallskip}
 & & Mar00 & Apr01 & Nov01 & Mar00 & Apr01 & Nov01 \cr
\hline
4918.247&$c~^2D_{3/2} - a~^2F_{5/2}$&13.60 &8.62 &11.50 &124.8 &61.5 &105.6  \cr
4984.205&$c~^2D_{3/2} - a~^2F_{7/2}$&2.64  &1.32 &2.00  &23.3  &11.7 &17.7   \cr
6153.623&$c~^2D_{3/2} - a~^2D_{3/2}$&      &     &3.76  &      &     &24.0  \cr
                            &       &                  &     &    &   &   & \cr
4927.288&$c~^2D_{5/2} - a~^2F_{7/2}$&14.70 &13.2b&13.40 &134.2 &120.5&122.4\cr
6079.567&$c~^2D_{5/2} - a~^2D_{5/2}$&      &     &2.86  &      &     &16.1   \cr
                            &       &               &   &     &    &       & \cr
6126.311&$b~^2F_{5/2} - a~^2F_{5/2}$&      &     &11.8  &      &     &65.5   \cr
6228.990&$b~^2F_{5/2} - a~^2F_{7/2}$&      &     &2.03  &      &     &10.9   \cr
                            &       &                   &    &     &   &  & \cr
6148.896&$b~^2F_{7/2} - a~^2F_{5/2}$&      &     &3.20  &      &     &17.6   \cr
6252.341&$b~^2F_{7/2} - a~^2F_{7/2}$&      &     &18.8  &      &     &100.3  \cr
                            &       &      &     &      &      &     &       \cr
6652.610&$b~^2G_{9/2} - a~^4F_{7/2}$&3.36 &1.97 &2.32   &15.9  & 9.3 &11.0  \cr
6727.670&$b~^2G_{9/2} - a~^4F_{9/2}$&5.12 &4.78 &5.38   &23.7  &22.1 &24.9   \cr
9652.716&$b~^2G_{9/2} - a~^2F_{7/2}$&0.71 &     &1.39   &1.7   &     &3.3   \cr
9408.356&$b~^2G_{9/2} - a~^2F_{5/2}$&2.62 &     &1.58   &6.3   &     &3.8    \cr
        &       &       &               &       &       &       &       \cr
6648.980&$b~^2G_{7/2} - a~^4F_{7/2}$&1.87 &1.67 &1.42   & 8.9  &7.9  &6.7   \cr
9645.089&$b~^2G_{7/2} - a~^2F_{7/2}$&1.02 &     &0.60   &2.4   &     &1.4    \cr
        &              &                  &       &       &      &       & \cr
7918.452&$b~^2D_{3/2} - a~^4F_{3/2}$&     &     &2.62   &      &     & 8.5   \cr
7977.897&$b~^2D_{3/2} - a~^4F_{5/2}$&     &     &6.03b  &      &     &19.3   \cr
8531.906&$b~^2D_{3/2} - b~^4F_{3/2}$&     &     &11.00  &      &     &30.5   \cr
8587.494&$b~^2D_{3/2} - b~^4F_{5/2}$&     &     &5.78   &      &     &15.8   \cr
8664.452&$b~^2D_{3/2} - b~^4F_{7/2}$&     &     &53.8b  &      &     &      \cr
 &      &       &       &                               &      &     &       \cr
7979.330&$b~^2D_{5/2}   - a~^4F_{7/2}$&   &     &6.03b  &     &     &        \cr
8568.400&$b~^2D_{5/2}   - b~^4F_{7/2}$&   &     &6.44   &      &     &  17.7\cr
8663.807&$b~^2D_{5/2}   - b~^4F_{9/2}$&   &     &53.80b &      &      &150.2 \cr
        &       &                       &       &  &       &              \cr
8335.590&$b~^2P_{3/2}   - a~^2F_{5/2}$&   &     &3.74   &     &     &    10.9\cr
8526.997&$b~^2P_{3/2}   - a~^2F_{7/2}$&   &     &13.90  &     &     &   38.6 \cr
 &      &       &                      &       &       &  &               \cr
8412.354&$b~^2P_{1/2}  - a~^2F_{5/2}$&    &     &7.72   &     &     &   22.0 \cr
10069.74b&$b~^4P_{5/2} - a~^4F_{5/2}$&1.70&   &3.57   &3.2   &     &  8.3  \cr
&$b~^4P_{3/2} - a~^4F_{3/2}$&&   &&&     &  \cr
 &      &       &                      &       &       &  &               \cr
10128.770&$b~^4P_{1/2} - a~^4F_{3/2}$&5.82    &     &7.84  &13.6  &    & 18.3\cr
        &       &                       &               &       &       &       &               \cr
\hline
\end{tabular}
\end{table*}

\begin{figure}
  \leavevmode
    \epsfysize=11cm
    \epsfbox{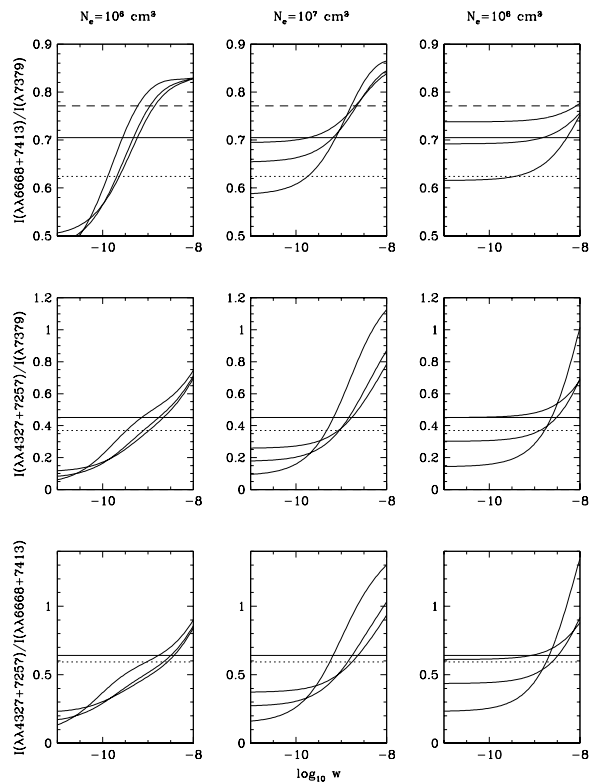}
\caption{[\ni2] emissivity line ratios
$I([\lambda\lambda6668+7413])/I([\lambda 7379])$,
$I([\lambda\lambda4327+7257])/I([\lambda 7379])$, and
$I([\lambda\lambda4327+7257])/I([\lambda\lambda 6668+7413])$
against the logarithm of the dilution factor. The line ratios
computed at $N_e=10^6$, $N_e=10^7$, $N_e=10^8$ \cm3 are presented
in separate panels located from left to right in this figure.
For each value of $N_e$ the line ratios are computed for
temperatures of 5000, 7000, and 9000~K.
The observed line ratios
are indicted by horizontal lines as dashed-line: mar00; dotted-line
: apr01; solid line: nov01}
\end{figure}

\subsection{[\ni2] diagnostics}

The combined contributions of collisional excitation and fluorescence 
make spectral diagnostics much more difficult than with traditional
collisionally excited lines. Most line ratios of interest 
become simultaneously 
sensitive to electron density, temperature, and the 
continuum radiation field (i.e. dilution factor), thus many lines have
to be studied as a whole to derive reliable conclusions about the 
physical conditions of the emitting plasma.

\begin{figure}
  \leavevmode
    \epsfysize=11cm
    \epsfbox{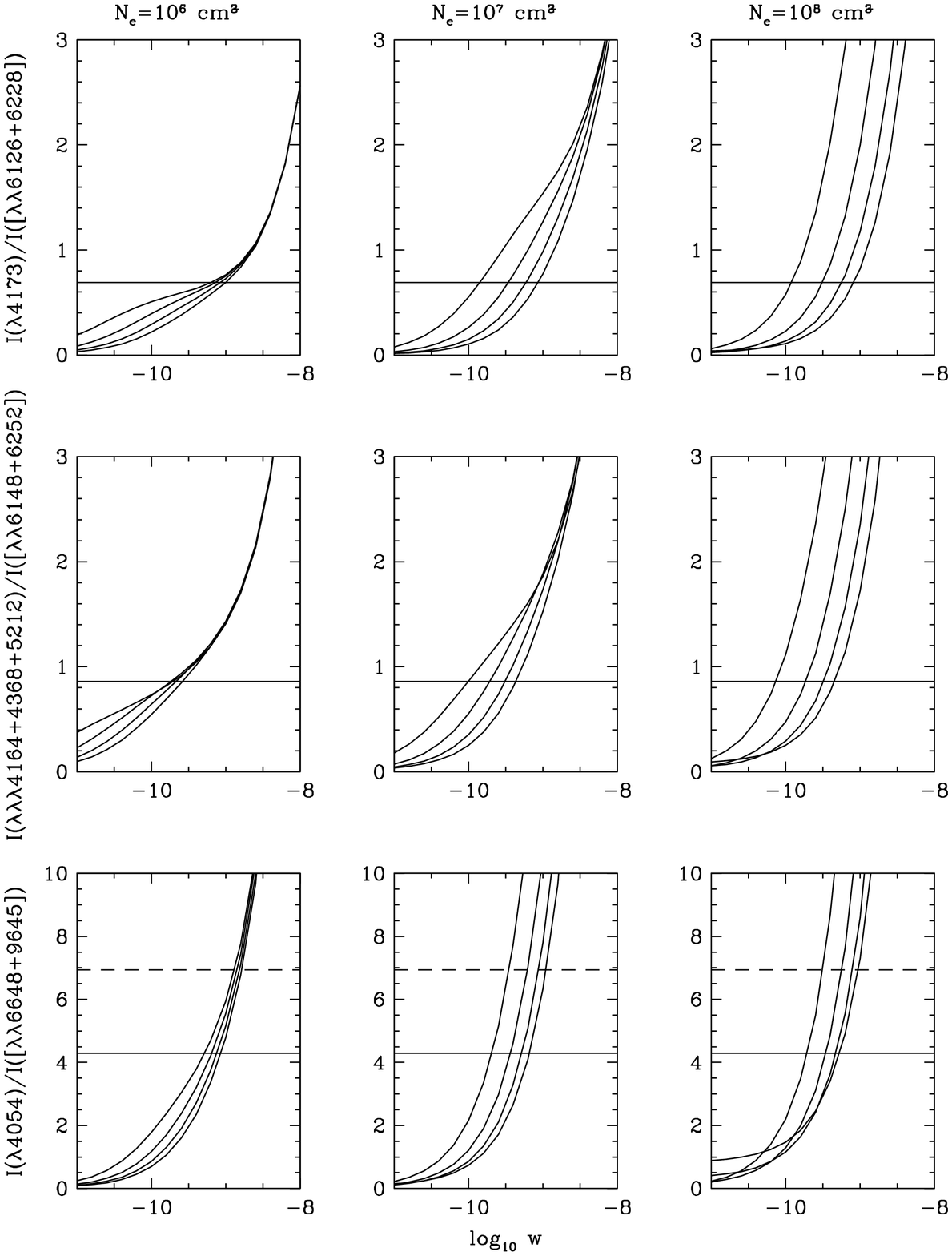}
\caption{\ti2 line ratios
between dipole allowed and forbidden line
against the logarithm of the dilution factor. The line ratios
computed at $N_e=10^6$, $N_e=10^7$, $N_e=10^8$ \cm3 are presented
in separate panels located from left to right in this figure.
For each value of $N_e$ the line ratios are computed for
temperatures of 4000, 5000, 6000, and 7000~K.
The observed line ratios for each observation
are indicted by horizontal lines as dashed-line: mar00; 
solid line: nov01}
\end{figure}

In the case of [\ni2] lines, only a few diagnostic line ratios can be used  
because
out of the seven dipole forbidden 
lines measured in the spectrum several share 
the same upper level. 
Fig.~4 shows the 
line ratio $I([\lambda\lambda6668+7413])/I([\lambda 7379])$,
$I([\lambda\lambda4327+7257])/I([\lambda 7379])$, and
$I([\lambda\lambda4327+7257])/I([\lambda\lambda 6668+7413])$
 against 
the logarithm of the dilution factor. The ratios were computed
for three values of the electron density, $N_e=$ $10^6$, $10^7$, and $10^8~cm^{-3}$
displayed in separated panels from left to right in the figure. In each panel
we present the line ratios as computed for temperatures of 5000, 7000, and 9000~K.
The observed lines ratios are indicated in the plots by horizontal lines
for each  observation.
Two conclusions can be derived from these plots in regards to the 
dilution factor in the filament. First, for the nov01 observation
the value of $log_{10}~w$ is in between -9.2 and -8.5. 
There is evidence for different dilution factors for each of the
three observations, with a difference as large as 1~dex between the 
mar00 and the apr01 observations, with the higher value
corresponding to the mar00 observation. The dilution factor from the
nov01 observation lies between the values for the
other two observations.
The spread in line ratios, and consequently on derived physical conditions, in 
the three observations are the result of different orientations
of the spectrograph slit with respect to the star (see Hartman et al. 2004). 
For the mar00 observations the slit was centered on \etacar with the 
fiducial, F2, blocking the starlight. This provided a spatial sampling of the 
Sr-emitting region extending radially from the central source. That this 
spectrum yields the highest value for the dilution factor 
is due to finding the radial distance for which both apply. 
The nov01 spectrum was taken at position angle -130 degrees and 
provided a slice of the Sr-filament that is tangential to the distance from 
the central source. As the position of the slit is more distant than the 
optimal position for highest dilution factor, we get a lower value for the 
dilution factor along this slit.
The dilution factor changes by an order of magnitude between the optimal 
radial position and the tangential sample because these observations sample 
slightly different spatial regions. Photons with energies below 7.9~eV
encounter 
increasing optical depths with distance from the central source. 
For these energies the
 main source of opacity is free-free opacity
(Bremsstrahlung). Assuming an electron temperature of the order
of $10^4$~K, density $\sim 10^7$ \cm3, and a difference in the path length
within the filament between the two observations
of $\sim~0".3~\approx~10^{16}$~cm, one gets an optical depth, $\tau$
$\approx 2$, which is consistent with the observed variation
in the dilution factor.

\subsection{[\ti2] diagnostics}

The large number of \ti2 and [\ti2] lines present in the
spectra allows us to constrain the physical conditions
of the filament

As discussed earlier, 
the combined contributions of collisional excitation and fluorescence
lead to complex temperature and density 
dependence in the relative strengths of dipole to  forbidden
emission lines. Nonetheless, such ratios are much more sensitive
to the strength of the dilution factor than to other parameters,
so they can be used to diagnose the value of the dilution factor.

In Fig.~5 we plot \ti2 line ratios  between dipole allowed and
forbidden lines against 
the dilution factor, together with the observed ratios. 
These line ratios are very sensitive to the value of the dilution factor, but 
depend relatively little on $T_e$ and $N_e$. 
From this figure, the line ratios for the nov01 observation
 are consistent with $log_{10}~w$ between -10.0 and -8.6, while the 
ratios from the mar00 spectrum indicate a higher dilution factor,
by as much as 0.5~dex. These results are in agreement with
those obtained with the [\ni2] lines. 

Considering all possible line ratios among forbidden lines we found one
that is primarily sensitive to the dilution factor, and depends little on
the temperature and density.
This ratio is shown in Fig.~6, as computed for $T_e$= 5000, 6000, and 7000~K
and $N_e=$ $10^6$, $10^7$, and $10^8$ \cm3. Once again the 
observed line ratios lay in the range of $log_{10}~w$ between -9.1 and -8.6.

\begin{figure}
  \leavevmode
    \epsfysize=11cm
    \epsfbox{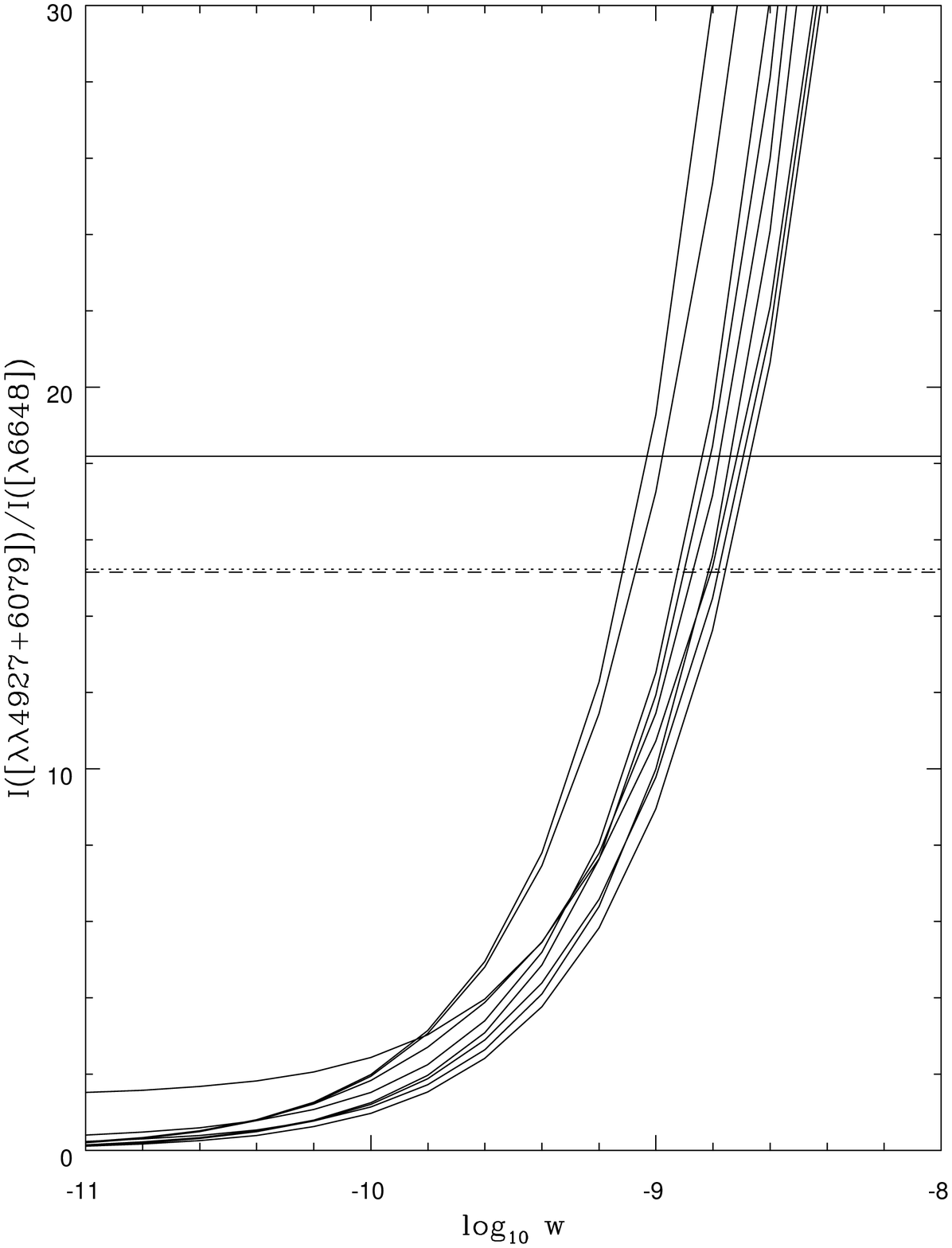}
\caption{\ti2 line ratios
between forbidden lines, 
$I([\lambda\lambda4927+6079])/I([\lambda6648])$, 
against the logarithm of the dilution factor. The line ratios
computed at $T_e=$ 4000, 5000, 6000, and 7000~K
and $N_e=10^6$, $N_e=10^7$, $N_e=10^8$ \cm3.
The observed line ratios for each observation
are indicted by horizontal lines as dashed-line: mar00; dotted-line: apr01; 
solid line: nov01}
\end{figure}

From the agreement between various \ni2 and \ti2 diagnostics we 
find the  value of  the dilution factor to be $log_{10}~w=-9.0\pm0.5$.
With the dilution factor constrained, one can diagnose 
other parameters of the emitting region.
Fig.~7 presents line ratios among \ti2 forbidden lines vs. the
electron temperature. The line ratios are shown in various panels 
for $N_e=$ $10^6$,
$10^7$, and $10^8$~\cm3. In each case we show the line ratios for
$log_{10}~w=-9.0$, -8.5, and -9.5. The observed line ratios from the 
nov01  spectrum are also indicated. Inspection of these plot
reveals that the best agreement for all the lines observed is obtained 
for $T_e=6000\pm 1000$~K.

\begin{figure}
  \leavevmode
    \epsfysize=11cm
    \epsfbox{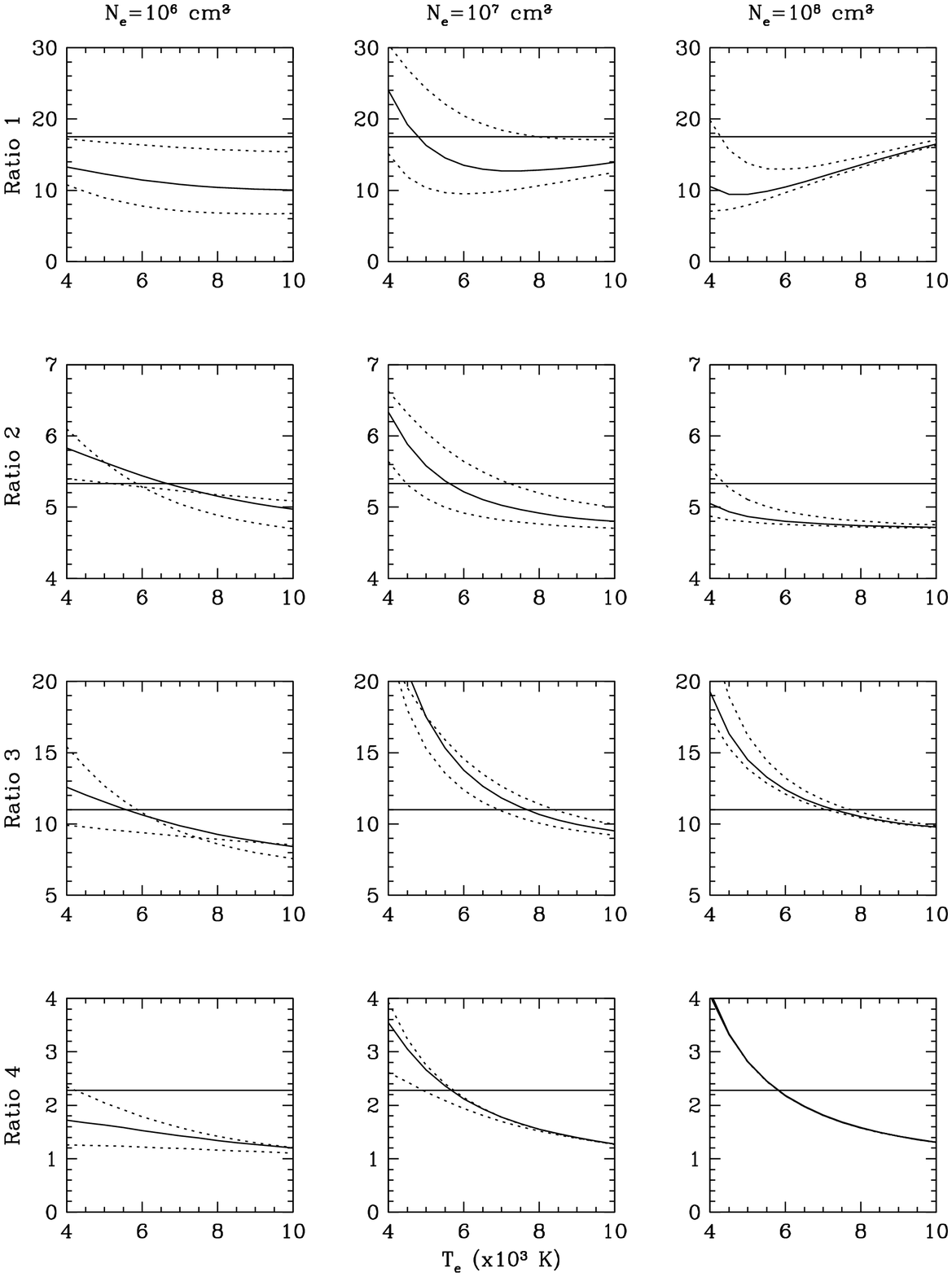} 
\caption{\ti2 line ratios
between forbidden lines vs. $T_e$. The line ratios are:
$I([\lambda\lambda6148+6252])/I([\lambda6648])$ (Ratio 1),
$I([\lambda\lambda6652+6727])/I([\lambda6648])$ (Ratio 2),
$I([\lambda\lambda7918+7977+8532+8587])/I([\lambda6648])$ (Ratio 3),
and $I([\lambda10069b])/I([\lambda6648])$ (Ratio 4).     
The line ratios
computed for 
$N_e=10^6$, $N_e=10^7$, $N_e=10^8$ \cm3 are displayed in separated 
panels from left to right. In each case $log_{10}~w$=-9.0 is indicated
by the solid line and $log_{10}~w$=-8.5 and -9.5 are indicated by 
dotted lines.
The observed line ratios for the nov01 are marked with 
horizontal lines}  
\end{figure}

\begin{figure}
  \leavevmode
    \epsfysize=11cm
    \epsfbox{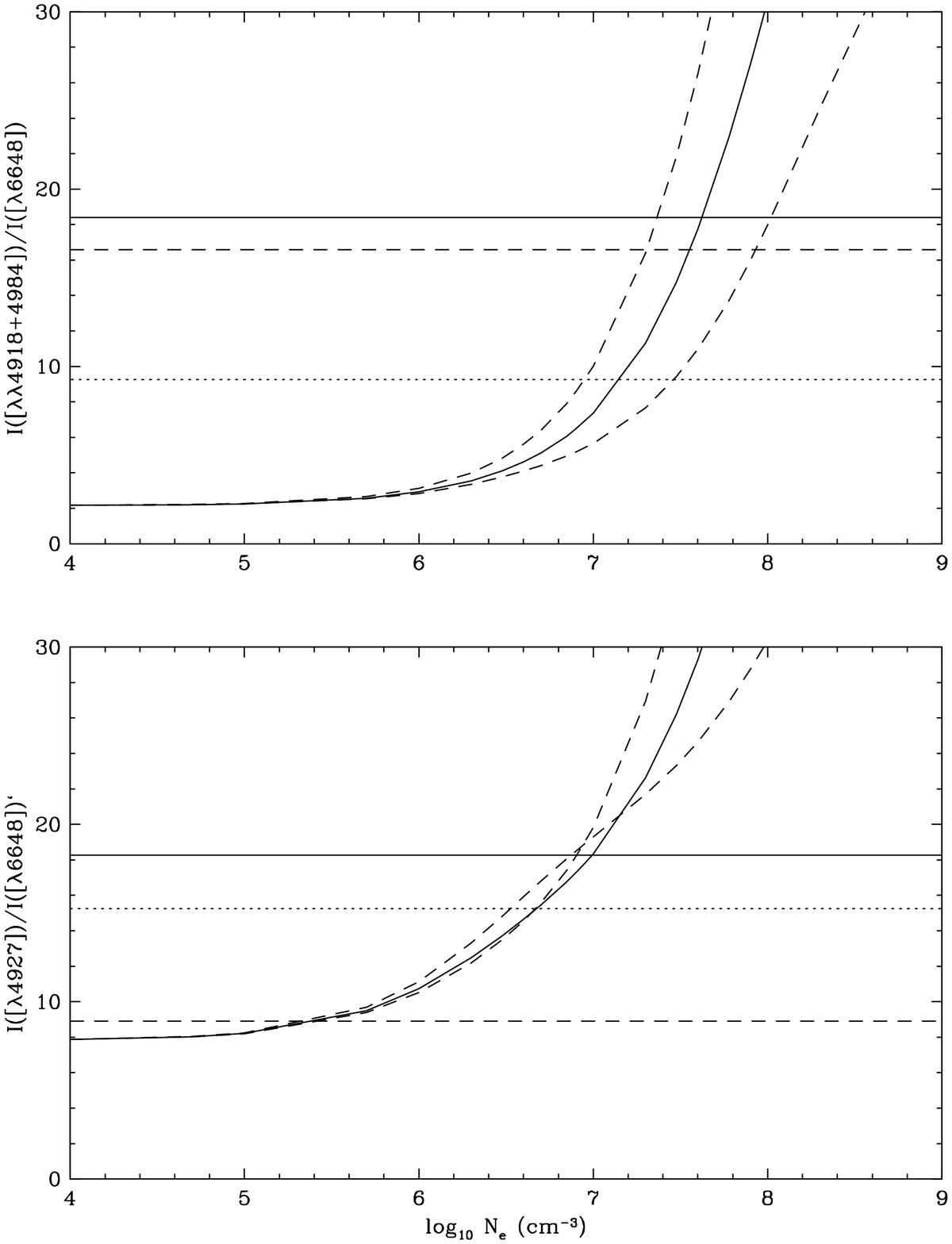}
\caption{\ti2 line ratios
between forbidden lines vs. $N_e$. 
The line ratios are
computed for
$T_e=6000$~K (solid line),  and $T_e=5000$~K and $T_e=7000$~K (dashed lines).
The dilution factor was fixed in $log_{10}~w$=-9.0. 
The observed line ratios 
are indicted by horizontal lines as dashed-line: mar00; dotted-line
: apr01; solid line: nov01}
\end{figure}

This relatively low temperature is consistent with the observed ionization
structure of Fe and Ni in the filament. The observed iron emission
arises mostly from neutral iron, indicating that this dominates
over \fe2. Because of the iron shielding of the filament 
it is expected that there would be few photons capable of ionizing Fe~{\sc i}.
But, even in the absence of photoionization collisional ionization
drives  most of the iron into \fe2
whenever electron temperatures exceed $\sim7000$~K (see
Arnaud and Raymond 1992). Below this temperature
neutral iron may be the dominant ionization stage. On the other hand,
under collisional ionization conditions 
the fraction of ionized to neutral nickel is more than twice the fraction
for iron at the same temperature. Thus, \ni2 is expected to remain the
dominant ionization stage for temperatures as low as $\sim 5000$~K.

Finally, we can determine 
the electron density. Fig.~8 depicts four line ratios as function
of the electron density for a fixed $log_{10}~w$=-9.0 and temperatures
of $6000\pm1000$~K. The electron density is readily diagnosed at 
around $10^7$~\cm3.

Given the physical conditions diagnosed above we compare the overall
agreement between the computed spectrum and observations. This 
comparison for the nov01 spectrum is shown in Table~7. 
The agreement is reasonably good considering the remaining uncertainties
in the atomic data and the shape of the pumping radiation continuum.

\subsection{Spatial variations of the dilution factor}

The \ti2 $I([\lambda\lambda6652+6727])/I([\lambda 4918])$ ratio is large across most of the 
spectral slit, which allows us to further investigate the spatial variations of the
dilution factor near the Sr-filament. To do this we integrated the
intensity over 0.13" wide regions of the spectrum from mar00       
and nov01.
The measured line ratios, corrected for extinction, and the derived
dilution factors from them are shown in Fig.~9.
The mar00 observations, when the F2 fiducial (0.85" long) 
was centered on \etacar with the large aperture extending radially 
from the star at the middle of the filament (1.3" from \etacar), clearly show
a linear decay of the dilution factor with distance from the central source.
Consistently with this picture, the nov01 slit that was
centered on the Sr-filament and perpendicular to the radial direction from the central source,
shows a maximum of the dilution factor at the middle of the filament and symmetric drop offs toward 
both sides of this position. 

\begin{figure}
  \leavevmode
    \epsfysize=11cm
    \epsfbox{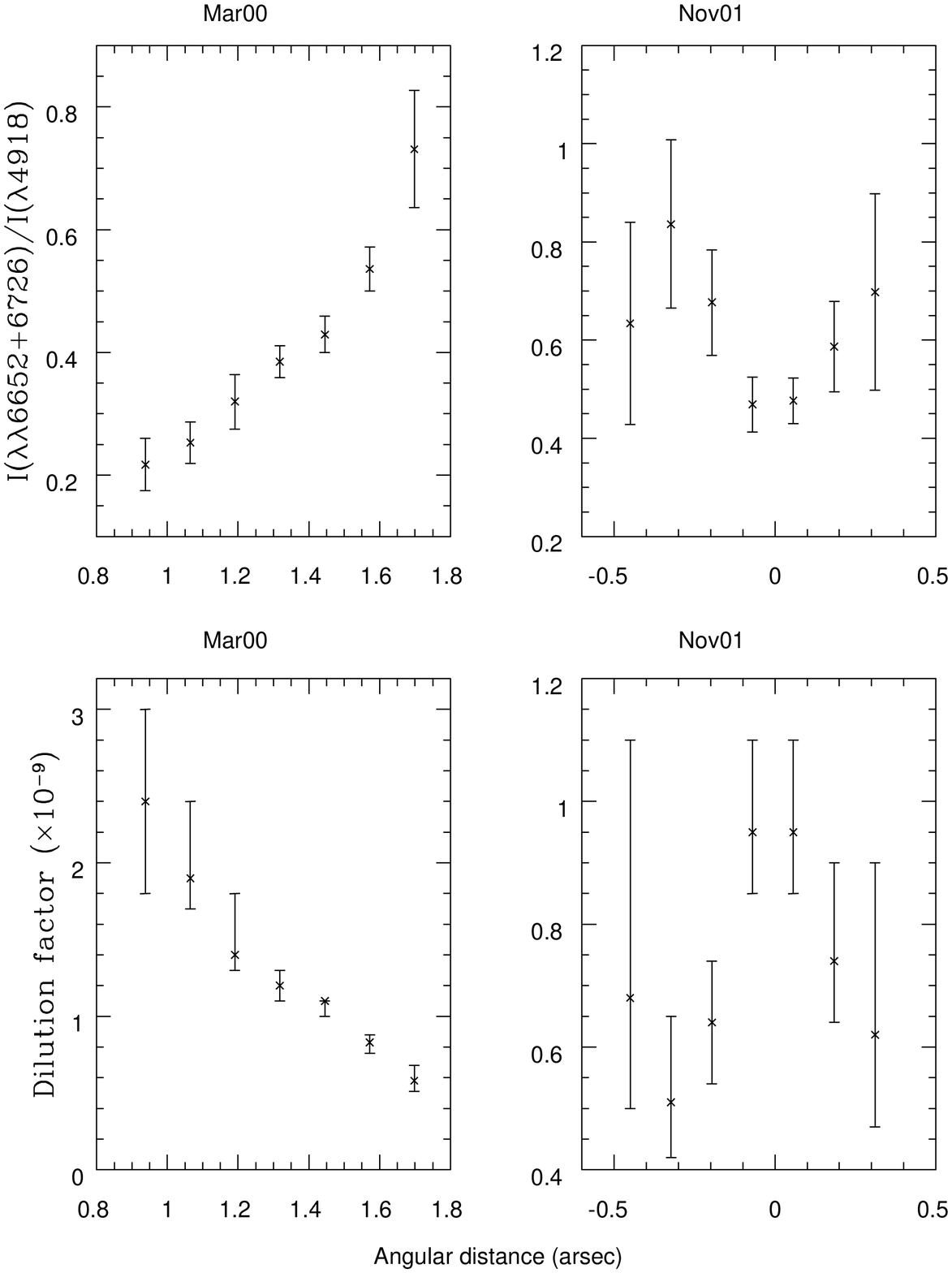} 
\caption{[\ti2] flux line ratios, 
$I(\lambda\lambda6652+6727)/I(\lambda 4918)$ (upper plots) and 
derived dilution factors (lower plots)
against the angular distance along the spectral slits of mar00 
 (radial across the filament
from the central source) and 
nov01 (tangential across the filament).} 
\end{figure}

\begin{table}
 \centering
\caption{Comparison between observed and theoretical line intensities
for the nov01 spectrum. Line intensities are relative to
$I([\lambda6648]$. The theoretical spectrum was computed for
$log_{10}=-9.0$, $T_e=6000$~K, and $N_e=10^7$~\cm3} 
\begin{tabular}{rrr}
\hline
\hline
$\lambda$   & \multicolumn{2}{c}{$I(\lambda)/I([\lambda6648])$} \cr
  &  Obs.  & Theo. \cr
\hline 
4918.26&         15.7&        6.38\cr  
4927.29&         18.2&        18.3\cr    
4984.20&          2.6&        1.01\cr  
6079.57&          2.4&        0.91\cr    
6126.31&          9.7&        7.23\cr     
6148.90&          2.6&        1.78\cr  
6153.62&          3.1&        0.25\cr 
6228.99&          1.6&        0.91\cr       
6252.34&         14.9&       11.70\cr 
6648.98&          1.0&        1.0\cr     
6652.61&          1.6&        1.49\cr     
6727.67&          3.7&        3.72\cr    
7918.45&          1.3&        0.51\cr    
7977.90&          2.9&        0.84\cr 
8335.59&          1.6&        0.80\cr    
8412.35&          3.3&        2.46\cr    
8527.00&          5.7&        4.50\cr    
8531.91&          4.5&        4.39\cr 
8568.40&          2.6&        8.98\cr   
8587.49&          2.4&        8.04\cr           
9645.09&          0.2&        0.98\cr  
9652.72&          0.5&        0.74\cr       
10069~b&          2.2&        2.12\cr  
10128.77&         2.7&        3.26\cr
\hline
\end{tabular}
\end{table}

\section{The titanium and nickel abundances in the Sr-filament}

\begin{figure}
  \leavevmode
    \epsfysize=10cm
    \epsfbox{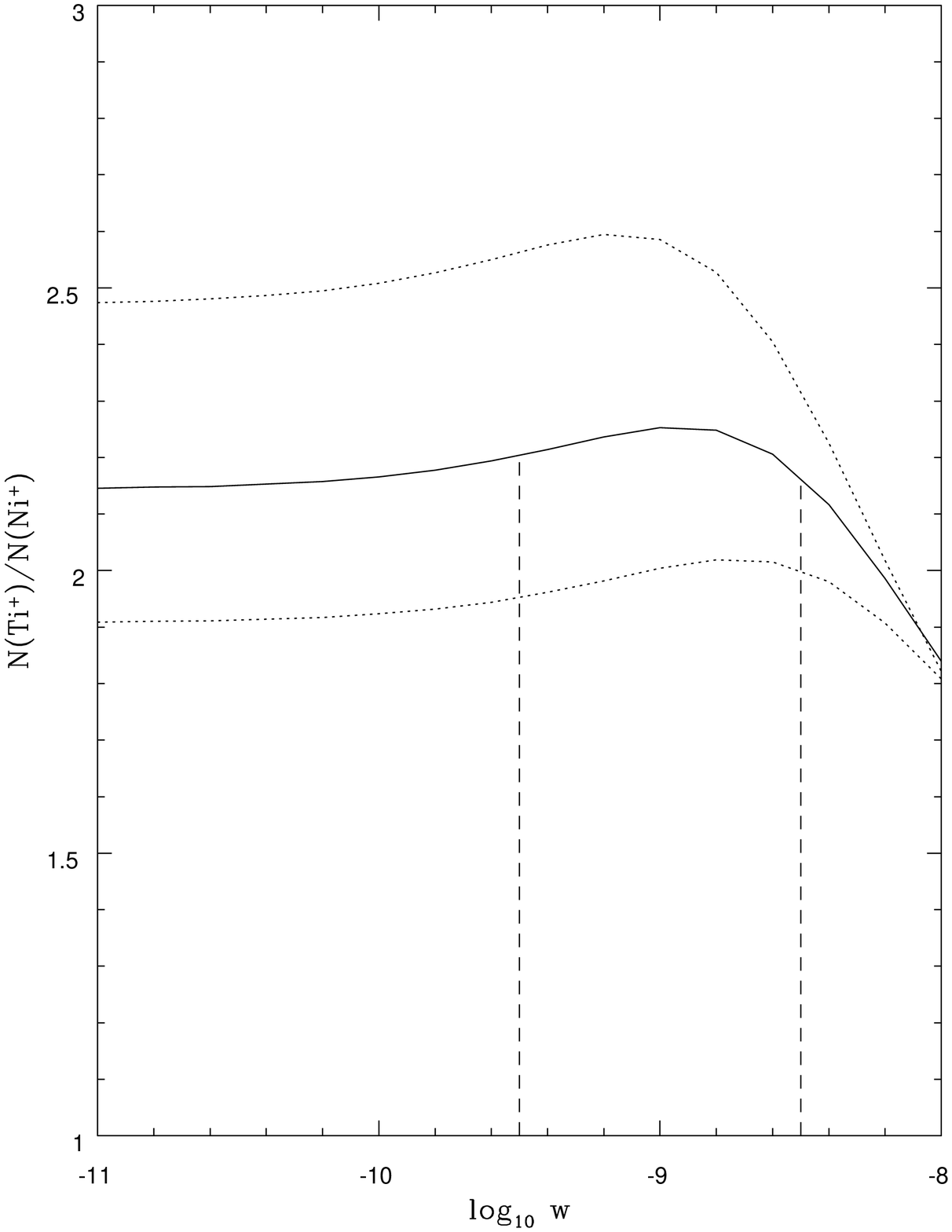}
\caption{$N(Ti^+)/N(Ni^+)$ abundance ratio from the spectrum of the 
Sr-filament as obtained for any given dilution factor between 
$10^{-11}$ and $10^{-9}$. The solid line depicts the abundance 
ratios for an adopted temperature of 6000~K, while the results
for temperatures of 5000 and 7000~K are indicated by the dotted lines.
The vertical dashed lines indicate the range of the dilution factor
as diagnosed from the spectrum}
\end{figure}

Once the physical conditions of the Sr-filament have been determined we proceed
to compute the relative abundances of \ni2 and \ti2. We use the nov01
spectrum because: (1) much  more spectral interval was covered during
that visit and the greatly increased number of lines, which permits a better 
statistical determination, and (2) the long aperture likely sampled similar 
regions for the \ni2 and \ti2 due to the tangential slit orientation.
Nonetheless, we have confirmed that 
the computed relative abundances from all other observations also agree, within
the uncertainties.

Fig.~10 depicts the $N(\ti2)/N(\ni2)$ as derived from the [\ti2] $\lambda 6648$
and [\ni2] $\lambda7379$ lines. The calculated abundance ratio is
plotted against the assumed value for the dilution factor and for
temperatures of $6000\pm1000$~K as diagnosed from the spectrum.
Given the fact that the lines observed from both species have
similar excitation energies and arise from levels with the 
same multiplicity and parity, they respond similarly to      
changes in the dilution factor. So we see that the uncertainty in the
abundance ratio is negligible for the mean temperature of 6000~K,
and uncertainty in the derived abundance does not exceed 20\% within the whole range of dilution parameters 
and temperatures. Thus, we conclude that 
$N(\ti2)/N(\ni2)\approx 2$, with a 
conservative uncertainty that may not exceed a  factor of three, including the dispersion
from all the lines which contains non-systematic uncertainties
in the atomic data and the treatment of the pumping continuum field.

To estimate the total $N(Ti)/N(Ni)$ abundance ratio we notice that 
only neutral and singly ionized ions are 
present in the spectra. 
It is expected that the ionic ratios Ti~{\sc i}/Ti~{\sc ii}
and Ni~{\sc i}/Ni~{\sc ii} are  similar since Ti~{\sc i} and 
Ni~{\sc i} have approximately the same ionization potential. 
Moreover, from observations one can conclude that Ni and Ti are both
singly ionized, as there are no  lines from the neutral stages of these
ions in the spectra. 
Thus, $N(Ti)/N(Ni) \approx N(Ti^+)/N(Ni^+) \approx 2$ in gas phase. 
This value is significantly higher than the best solar abundance ratio of $\sim$0.05.
In other words, the gas phase $N(Ti)/N(Ni)$ abundance ratio in the Sr-filament 
is nearly two orders of magnitude greater than solar value.

Further, we have checked on the $N(\ni2)/N(\fe2)$ abundance ratio in the filament.
This calculation is complicated by the fact that the [\fe2] emission comes from
a more extended region in velocity space. Thus, we 
only consider the integrated flux of [\fe2]
lines within the range in velocity space that coincides the [\ni2] emission we
find a value for the $N(\ni2)/N(\fe2)$ ratio roughly consistent with the solar 
$N(Ni)/N(Fe)$ ratio, assuming that 
the ionic fractions of \ni2 and \fe2 are 
tied to each other by charge exchange and are thus within a factor of two of each other.
It must be noted, however, that the filament exhibits prominent emission
from Fe~{\sc i}, suggesting a large abundance of this ion, but no emission
from Ni~{\sc i} or Ti~{\sc i} within the sensitivity of the observations. 

\section{Titanium, nickel, and iron condensation in \etacar}

\begin{figure}
  \leavevmode
    \epsfysize= 9cm
    \epsfbox{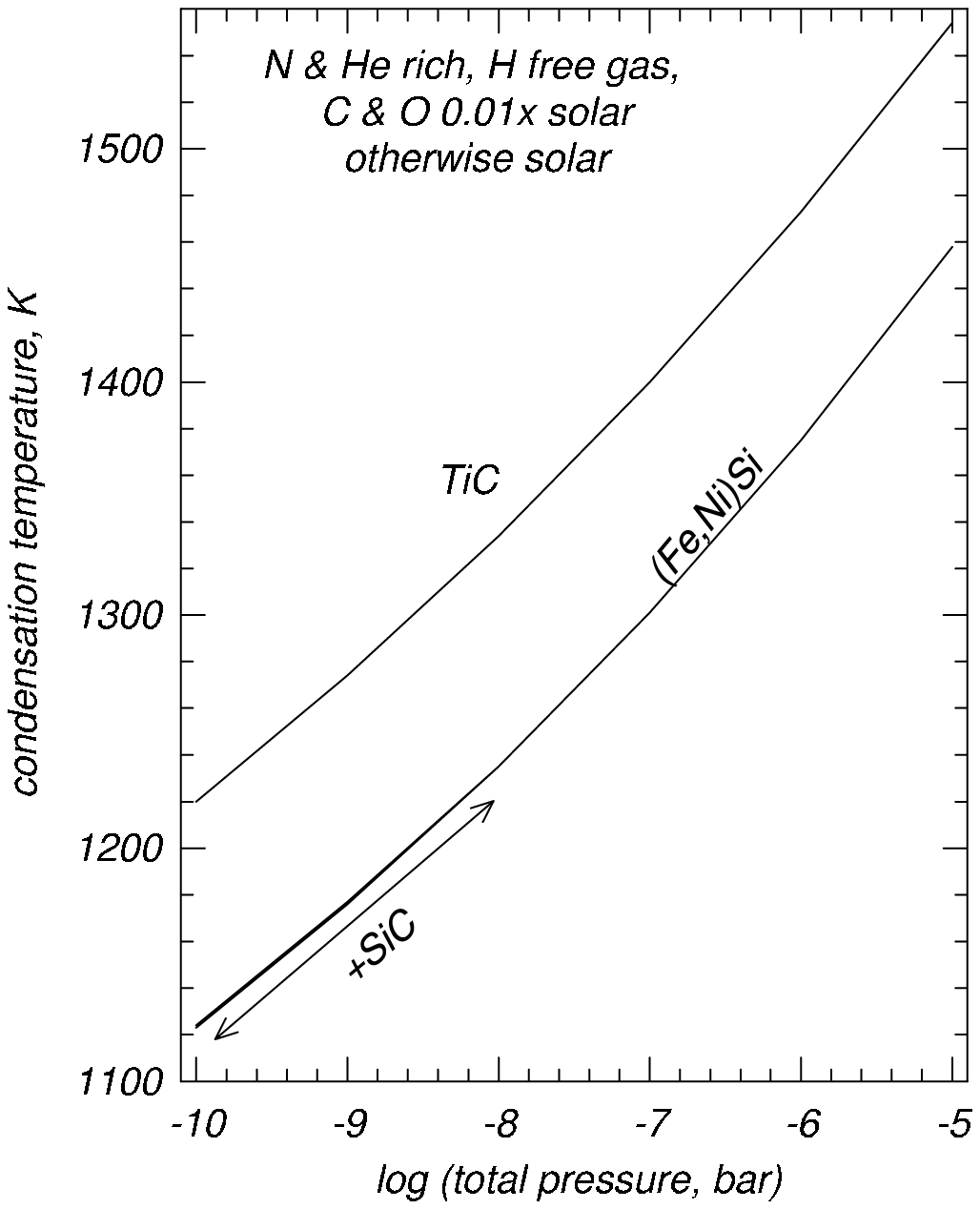}
\caption{Condensation temperatures as a function of total pressure} 
\end{figure}

\begin{figure}
  \leavevmode
    \epsfysize= 8cm
    \epsfbox{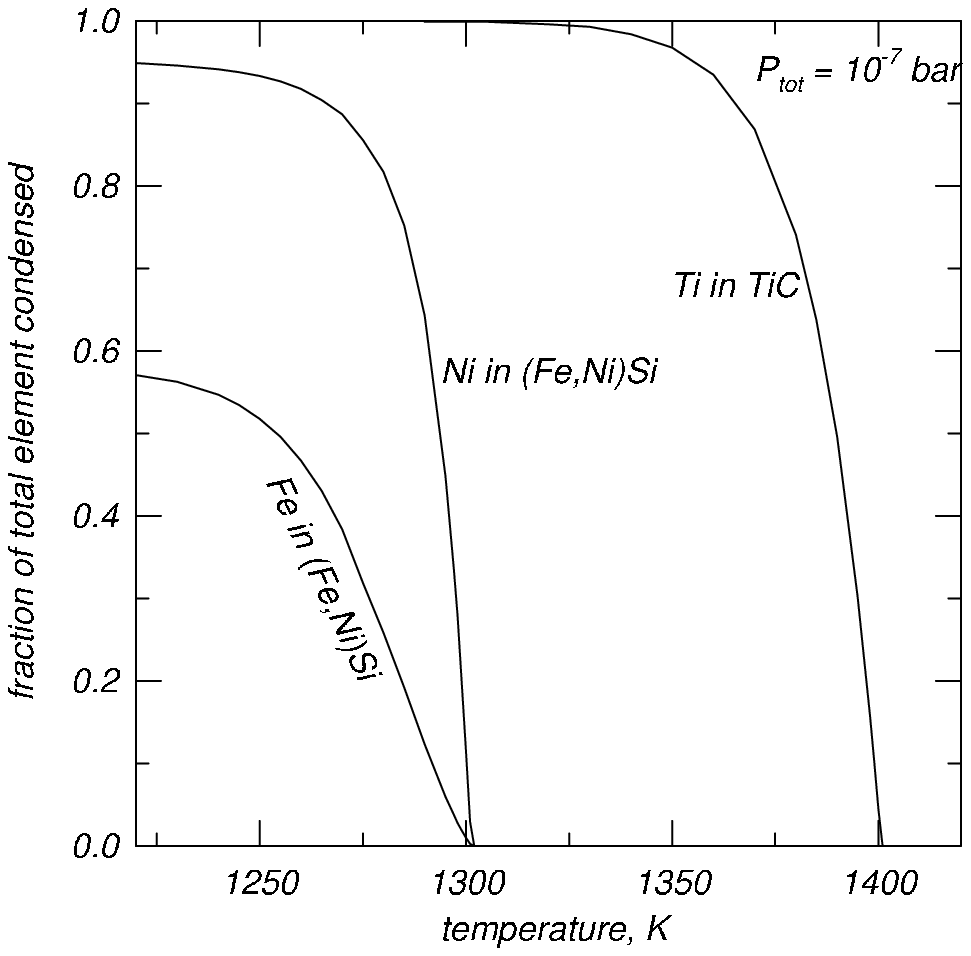}
\caption{The fraction of an element condensed as a function of temperature
at a constant total pressure of $10^{-7}$ bar}
\end{figure}

The observed large overabundance of titanium with respect to both nickel and iron in 
the gas phase is quite striking. Ti/Fe abundance ratios higher than solar by up to 0.4 dex 
are known to occur in metal-poor and extremely metal-poor stars (Ryan et al. 1996), but 
to the best of our knowledge there are no reports of such a high overabundance as we 
have analyzed here. 
It is likely that the abnormal relative abundances found here are only representative of the gas phase, rather than the total absolute abundances.
Under the conditions of the Sr-filament nearly 90\% of Fe and Ni are condensed into grains. Thus, a large Ti/Ni gas phase abundance ratio would be attained if some kind of destruction mechanism of Ti-bearing condensates acted in the filament. Although we do not have a clear explanation for such a mechanism we devote this section of the paper to study the chemistry of the Ti, Ni, and Fe under the conditions of the Sr-filament.

Condensation temperatures of Ti, Fe, and Ni are 
somewhat different and Ti is generally more refractory than Fe and Ni during 
condensation from a solar composition gas (e.g., Lodders 2003). As described below, Ti 
also forms a more refractory condensate than Fe and Ni from the overall composition 
relevant for the ejecta here. 
Condensation temperatures depend on bulk composition and total pressure. In a solar 
composition gas with C/O = 0.5, titanium condenses as calcium titanate (e.g., perovskite, 
CaTiO$_3$), followed by an iron and nickel metal alloy at somewhat lower temperatures 
(Lodders 2003). Trace elements such as Sr condense into solid solution with major element host 
phases, and condensation of Sr as "SrTiO$_3$" into CaTiO$_3$ proceeds at temperatures 
between that of CaTiO$_3$ and metal alloy formation.
As long as the overall composition of \etacar's ejecta remained near-solar, solar-type 
condensates should also have formed. 
However, 
in the Sr-filament and in the rest of the emission ejecta within the Homunculus, the 
Little Homunculus and the Weigelt Blobs, very little oxygen has ever been detected. 
Moreover, in the Weigelt Blobs where [O~{\sc i}] $\lambda6300$ emission is seen, 
oxygen seems to be underabundant by at least two orders of magnitude (e.g. Thackeray 
1967, Zethson 2001, Verner et al. 2005). 
In the outer ejecta carbon and oxygen are severely depleted 
relative to solar abundances (Davidson et al. 1982; Dufour et al 1997; Smith and Morse 2004). Hydrogen is also depleted 
whereas N and He are enriched. In order to explore the condensation chemistry, we adopt 
a solar composition gas in which solar C and O abundances are both reduced by a factor 
of 100. The N abundance is increased by the amount removed from the solar C and O 
abundances so that the sum of solar CNO nuclei is preserved. The He abundance is taken 
as solar plus a quarter of the solar H abundance, and the H abundance is set to zero. Note that the chemistry of Ti, Fe, Ni, and Si is relatively insensitive to the total H abundance because there are no major H-bearing gases for those elements. 
Even if H were fairly abundant, the following condensation results would change little.  
Descriptions of the thermochemical computation methods are given in Lodders (2003) and Lodders \& Fegley (1997).
The results of this calculation serve as a good illustration of the dust
chemistry, and should remain qualitatively correct despite variations
of the precise relative abundances. 
The condensates (Fig.~11) for this "CNO" processed composition are first TiC, and at 
about 100~K lower temperature, an iron-nickel silicide alloy, (Fe,Ni)Si. 
Below total 
pressures of $\sim 10^{-7}$ bar, SiC condensation temperatures are within a few degrees of those 
of the silicide alloy.

The condensates from the "CNO" processed composition are the same as appear when 
C and O abundances of an otherwise solar composition gas are only slightly (i.e. orders 
of factor 2) modified so that C/O $\ge$ 1 (Lodders \& Fegley 1995, 1999). Here the C/O 
ratio remains solar but the large decrease in absolute C and O abundances is responsible 
for the appearance of the carbides and silicide because there is no oxygen for SiO and 
silicate or oxide formation. 

The elemental composition here has enough C for condensation of all Ti as TiC (the 
total Ti/C is $\sim$ 0.03) but only a little SiC can form because Si is more abundant than C 
(Si/C $\sim$ 14) and Si condenses also into (Fe,Ni)Si. The overall C abundance is so low that 
graphite does not condense at the relatively high temperatures where TiC and (Fe,Ni)Si 
form. If carbon were completely absent here, titanium nitride (TiN, osbornite) would 
condense instead of TiC at about the same temperature, SiC would be completely absent, 
and condensation of (Fe,Ni)Si would remain unchanged.

Condensation 
temperatures describe when a compound starts to form (i.e., where the saturation ratio 
reaches unity) but lower temperatures are needed to fully remove an element from the gas 
into a condensate, as shown for a constant total pressure in Fig.~12. Complete titanium 
removal from the gas requires a drop of about 100~K below the condensation temperature 
of TiC and all Ti is condensed at temperatures when (Fe,Ni)Si condensation begins to 
proceed. 

Given the nature of the Ti-, Ni-, and Fe-bearings described here we suggest that
selective photoevaporation of Ti-bearings is a possible explanation for the gas phase Ti/Ni abundance found in the filament. Two different scenarios may be 
suggested: (1) there was some spatial separation between the TiC and(Ni,Fe)Si condensates, related to their different condensation temperatures, that left the
TiC rich gas more exposed to photoevaporation; (2) the TiC grains are 
generally smaller than the (Ni,Fe)Si grains, 
and so the TiC condensates were much less resistant to photoevaporation.
 
These scenarios sketched out here seem promising to explain the 
observations and need to be studied in more detail. Further, it is 
important to determine the abundance ratios of other elements of different volatility. 

\section{Summary  and conclusions}  

We have studied the \ni2 and \ti2 spectra of the Sr-filament in \etacar,
as taken in three different epochs with the HST. In doing this we build
spectral models for the two ions using the best published data for \ni2,
while we carry out extensive calculations of radiative and electron impact rates
for \ti2.

The spectral models were employed for diagnosing the physical conditions of the emitting region.
The results are generally consistent with previous determinations using [\sr2]
and \sr2 lines. Moreover, the [\ti2] lines allow us to constrain the conditions
for observations with the aperture across
the \etacar's central source 
to $N_e\approx 10^7$ \cm3, $T_e=6000\pm1000$~K, and dilution factor 
$log_{10}w=-9.0\pm0.5$ for an assumed blackbody continuum radiation with temperature of
35~000~K.
For observations with the slit perpendicular to the radial direction the
dilution factor seems to be somewhat lower. 

Study of the observed [\ti2] emission reveals 
a very large overabundance of Ti in gas phase with respect to Ni and Fe relative to the cosmic values. 

We study the chemistry of Ti, Ni, and Fe in CNO-cycled gas and find that Ti condenses as TiC, or alternatively as TiN, while Ni and Fe are locked into (Ni,Fe)Si. Moreover, the TiC condenses at higher temperatures than Ni- and Fe-bearings. Later, we suggest that the TiC would be selectively destroyed by the
stellar UV radiation leading to a relative overabundance of Ti in gas phase.

 This explanation for the unusual Ti/Ni ratios in the Sr-filament requires that significant dust and molecules had formed in
  this material shortly after the Great Eruption, but have since been
  largely photoevporated here.  In this context, it is very
  interesting that near-IR emission from molecular hydrogen is seen
  throughout the polar lobes of the Homunculus Nebula (Smith 2002a),
  but this same H$_2$ emission is not seen at comparable levels from
  the equatorial skirt in the vicinity of the Sr-filament (Smith 2002a,
  2004).  Today, the gas densities we infer in the Sr filament
  ($\sim$10$^7$ cm$^{-3}$) are very high compared to many other
  circumstellar nebulae around massive stars, and are similar to the
  extremely high densities seen in the walls of the Homunculus (e.g.,
  Gull et al. 2005; Nielsen et al. 2005).  A weaker radiation field at
  larger radii is not the only explanation for this difference: H$_2$
  emission is seen both in the polar caps of the Homunculus at larger
  radii from the star than the Sr-filament, and in the side walls of
  the polar lobes near the equator, even {\it closer} to the star than
  the Sr filament (Smith 2004).  In the past when the Sr-filament
  ejecta were even denser, it is conceivable that H$_2$ or other
  molecules may have existed, but have since been destroyed.  If true,
  this would seem to support the scenario of unusual chemistry
  resulting from selective photoevaporation and CNO processed ejecta,
  as we have suggested here.

\section*{Acknowledgments} 

The observations of the strontium filament were made with the Space
Telescope Imaging Spectrograph on the NASA/ESA Hubble Space Telescope
and were obtained by the Space Telescope Science Institute, which is
operated by the Association of Universities for Research in Astronomy,
Incorporated, under NASA contract NAS5-26555. Support for this research
was provided in part by NASA through the STIS GTO grant and grants from
the Space Telescope Science Institute. H. Hartman is supported by a grant
from the Swedish National Space Board. 
N. Smith was supported by
NASA through grant HF-01166.01A from STScI.
Work by KL supported in part by grant NNG04G157A from the NASA Astrobiology
Institute. 
Data for this analysis was gathered under GO programs headed by Kris Davidson and GO and GTO programs headed by Theodore Gull. 

We thank Sveneric Johansson for much guidance and advice.

\end{document}